\documentclass[twocolumn]{aastex631}

\usepackage{graphicx}	% Including figure files
\usepackage{amsmath}	% Advanced maths commands
\usepackage{multirow}
\usepackage{subcaption}
\usepackage{ulem}
\usepackage{natbib}

\def\hmpc{h^{-1}{\rm Mpc}}

\def\hmsun{{h^{-1} M_{\odot}}}

\def\astrid{\texttt{ASTRID} }

\graphicspath{{./}{figures/}}

\defcitealias{Chen2024}{MAGICS-I}
\defcitealias{Zhou2024MagicsII}{MAGICS-II}

\begin{document}

\title{MAGICS III. Seeds sink swiftly: nuclear star clusters dramatically accelerate seed black hole mergers}

\author[0009-0008-0384-3798]{Diptajyoti Mukherjee}
\affiliation{McWilliams Center for Astrophysics and Cosmology \\
Department of Physics, \\
Carnegie Mellon University, \\
Pittsburgh, PA 15213, USA}

\author{Yihao Zhou}
\affiliation{McWilliams Center for Astrophysics and Cosmology \\
Department of Physics, \\
Carnegie Mellon University, \\
Pittsburgh, PA 15213, USA}

\author{Nianyi Chen}
\affiliation{McWilliams Center for Astrophysics and Cosmology \\
Department of Physics, \\
Carnegie Mellon University, \\
Pittsburgh, PA 15213, USA}

\author{Ugo Niccolò Di Carlo }
\affiliation{SISSA - International School for Advanced Studies, \\
via Bonomea 365, I-34136 \\
Trieste, Italy}

\author{Tiziana Di Matteo}
\affiliation{McWilliams Center for Astrophysics and Cosmology \\
Department of Physics, \\
Carnegie Mellon University, \\
Pittsburgh, PA 15213, USA}



\begin{abstract}

Merger rate predictions of Massive Black Hole (MBH) seeds from large-scale cosmological simulations differ widely, with recent studies highlighting the challenge of low-mass MBH seeds failing to reach the galactic center, a phenomenon known as the seed sinking problem. In this work, we tackle this issue by integrating cosmological simulations and galaxy merger simulations from the MAGICS-I and MAGICS-II resimulation suites with high-resolution $N$-body simulations. Building on the findings of MAGICS-II, which showed that only MBH seeds embedded in stellar systems are able to sink to the center, we extend the investigation by incorporating nuclear star clusters (NSCs) into our models. Utilizing $N$-body resimulations with up to $10^7$ particles, we demonstrate that interactions between NSCs and their surrounding galactic environment, particularly tidal forces triggered by cluster interactions, significantly accelerate the sinking of MBHs to the galactic center. This process leads to the formation of a hard binary in $\lesssim 500$ Myr after the onset of a galaxy merger. Our results show that in 8 out of 12 models, the high stellar density of the surrounding NSCs enhances MBH hardening, facilitating gravitational wave (GW) mergers by redshift $z = 4$. We conclude that at $z > 4$, dense NSCs serve as the dominant channel for MBH seed mergers, producing a merger rate of $0.3$--$0.6\, \mathrm{yr}^{-1}$ at $z = 4$, which is approximately 300--600 times higher than in non-NSC environments. In contrast, in environments without NSCs, surrounding dark matter plays a more significant role in loss-cone scattering.

\end{abstract}



\section{Introduction} \label{sec:intro}
The formation and evolution of massive black holes (MBHs) stand as one of the most intriguing phenomena in astrophysics, playing a pivotal role in shaping the cosmic landscape from the early universe to the present epoch. Supermassive Black Holes (SMBHs) are quite ubiquitous, as observations suggest, and present in almost all galactic centers in local galaxies \citep[e.g.,][]{Miyoshi1995Natur.373..127M,Tremaine2002ApJ...574..740T,KormendyHo2013ARA&A..51..511K}. Intermediate Mass Black Holes (IMBHs) ranging from $10^3 M_{\odot}$ to $10^6 M_{\odot}$ have also been found to exist at the centers of several local dwarf galaxies \citep[e.g.,][]{Reines2013ApJ...775..116R,Greene2020ARA&A..58..257G}.  Observations, particularly from the James Webb Space Telescope (JWST), have unveiled a wide population of MBHs with masses $10^6 M_{\odot} - 10^8 M_{\odot}$ in the high redshift universe \citep[e.g.,][]{Larson2023ApJ...953L..29L,Ubler2023A&A...677A.145U,Kocevski2023ApJ...954L...4K,Harikane2023ApJ...959...39H,Goulding2023ApJ...955L..24G,Maiolino2023arXiv230801230M,Maiolino2024Natur.627...59M,Matthee2024ApJ...963..129M}. Scaling relations derived from local galaxies show that these MBHs are typically overmassive compared to their hosts \citep[][]{Pacucci2023ApJ...957L...3P,Goulding2023ApJ...955L..24G} and pose new challenges to our understanding of the formation and growth of MBHs in the high redshift universe.

In recent years, significant strides have been made in understanding the origins of MBHs, suggesting that they may originate from a seed population at $z\sim 20 - 30$ \citep{Barkana2001PhR...349..125B} which then grows in mass via mergers and accretion \citep[e.g.,][]{Dayal2019MNRAS.486.2336D,Pacucci2020ApJ...895...95P,Piana2021MNRAS.500.2146P,Bhowmick2024arXiv240203626B}. These seeds can form from the direct collapse of pristine gas clouds \citep[e.g.,][]{Begelman2006MNRAS.370..289B,Mayer2010Natur.466.1082M} or from the collapse of massive PopIII stars and runaway growth in dense stellar systems \citep[e.g.,][]{PortegiesZwart1999A&A...348..117P} such as nuclear star clusters (NSCs) \citep[e.g.,][]{Devecchi2009ApJ...694..302D,Lupi2014MNRAS.442.3616L,Das2021MNRAS.503.1051D,Askar2023arXiv231112118A,Kritos2023PhRvD.108h3012K} or young star clusters \citep[e.g.,][]{DiCarlo2019MNRAS.487.2947D,DiCarlo2021MNRAS.507.5132D}. However, reconciling theoretical predictions of different seeding models with observations remains a challenge due to the low masses of these seeds and their faint electromagnetic signatures \citep[][]{Pacucci2017ApJ...850L..42P}.

With the advent of gravitational wave (GW) astronomy using LIGO-Virgo interferometers \citep[e.g.,][]{abbott2017gw170814,abbott2020gw190412,abbott2020gw190425,abbott2020gw190814} and Pulsar Timing Array (PTA) \citep[][]{Mingarelli2017NatAs...1..886M,Kelley2018MNRAS.477..964K,nanograv2023ApJ...951L...8A,nanograv2023ApJ...951L..50A}, a new pathway has emerged to understand and constrain various seeding models in the early universe, if seed BHs can form MBH binaries and merge. In particular, MBH binaries with masses $10^4 M_{\odot} - 10^7 M_{\odot}$ are one of the main targets of the space based GW detectors like LISA \citep{amaro2017laser} or TianQin \citep{luo2016tianqin} which should be able to detect mergers of MBHs out to $z > 20$. However, the dynamics of these $10^3 M_{\odot} -10^6 M_{\odot}$ MBH seeds are notoriously difficult to model as they involve resolving processes that can bring them from kiloparsec scales to coalescence, producing a wide range in theoretical predictions for merger rates. 

The process leading to the GW coalescence of MBH binaries is often conceptualized as a three-step journey \citep[][]{Begelman1980Natur.287..307B,Merritt2013degn.book.....M}. Initially, the dynamical friction (DF) \citep{Chandra1943ApJ....97..255C} exerted by stars, dark matter (DM), and interstellar gas comes into play, reducing the angular momentum of the MBHs, leading them to sink to the center of the merged galaxy. When the two MBHs get close enough, they form a bound binary, marking the onset of the second stage. During this phase, the binary's separation reduces due to a combination of dynamical friction and three-body scattering events. In the penultimate stage, further orbital decay ensues, primarily driven by three-body scattering processes. However, if insufficient scattering interactions between MBH binaries and stars occur, the orbital decay of the binary can stall before the binary can coalesce via the emission of GWs \citep[e.g.,][]{Milo2003AIPC..686..201M,Vasilev2015ApJ...810...49V}.  This highlights the importance of the environment surrounding the MBH binary on its fate.

The self-consistent evolution of MBH seeds and their galactic environments is generally treated using cosmological simulations, which offer a comprehensive framework for understanding the intricate processes governing seed growth and mergers \citep[see][for a review]{Amaro2023LRR....26....2A}. However, large-volume cosmological simulations lack sufficient resolution at scales of $\lesssim \mathrm{kpc}$. Physical processes like dynamical friction are usually treated in a subgrid fashion \citep[][]{Tremmel2015MNRAS.451.1868T,Chen2022MNRAS.510..531C,Ma2023MNRAS.519.5543M,Damiano2024arXiv240312600D}  leading to uncertainties in the dynamics of MBH seeds, especially at separations of $\leq $kpc from the galactic centers. A number of recent studies have found that MBH seeds are
inefficient at sinking to the center of the galaxy leading to longer
merger timescales than previously expected \citep[][]{Pfister2019MNRAS.486..101P, Ma2021MNRAS.508.1973M, Partmann2023arXiv231008079P}, creating the so-called ``seed-sinking problem''. Crucially, large softening lengths and lower mass resolution in cosmological simulations prevent the resolution of dense stellar systems such as NSCs, which can play a major role in the formation and evolution of MBH seeds \citep[e.g.,][]{Devecchi2009ApJ...694..302D,Das2021MNRAS.503.1051D,Lupi2014MNRAS.442.3616L,Askar2021MNRAS.502.2682A}.

NSCs are some of the densest known stellar systems, and galaxies hosting both NSCs and MBHs at their centers are quite prevalent across galaxy masses and morphologies \citep{Neumayer2020A&ARv..28....4N}. Due to the large stellar density at their centers, they provide an ideal environment for the formation and growth of MBH seeds. Using semi-analytical models, \citet{Kritos2023PhRvD.108h3012K} find that NSCs in dwarf galaxies can provide suitable environments for the production of MBH seeds of masses up to $10^6 M_{\odot}$. In the densest and most compact NSCs, runaway growth and gas accretion can lead to formation of $10^6 M_{\odot}$ MBHs in $\sim 100$ Myr. 

\citet{Shi2024arXiv240517338S} perform multi-physics simulation and find that low mass MBH seeds present in star clusters hierarchically merge, leading to the formation of a proto-bulge or a dense NSC. The star clusters surrounding the seeds help them migrate efficiently to the center of the galaxy, resulting in a rapid formation of seed MBH binaries with separations $\leq 1$ pc. The process is much faster than the DF timescale of such MBH seeds, if they were isolated. On a similar note, recent works have found that MBHs embedded in NSCs undergo accelerated mergers\citep[][]{Ogiya2020MNRAS.493.3676O,Khan2021MNRAS.508.1174K,Mukherjee2023MNRAS.518.4801M, Chen2024}. The additional mass surrounding the MBHs helps them sink more efficiently to the galactic center. Once these MBHs reach a separation of $\lesssim 50 \mathrm{pc}$, tidal interactions of the NSCs help them sink from 20-50 pc to a few milliparsec in a span of $\sim$ Myr, leading to the formation of a hard binary faster than the DF timescale. %The surrounding medium also ensures a steady supply of stars on loss-cone orbits to further harden the binary

Recent studies have systematically explored favorable conditions under which MBH seeds can merge, leveraging multi-scale examinations. Large-volume cosmological simulations, such as \texttt{ASTRID} \citep{Bird-Astrid, Ni-Astrid,Chen2022MNRAS.514.2220C}, along with galaxy merger re-simulations like MAGICS-I \citep{Chen2024}, have demonstrated that galaxy mergers, particularly those involving dwarf galaxies, play a critical role in driving MBH seed mergers in the high-redshift universe. Building on MAGICS-I, MAGICS-II \citep{Zhou2024MagicsII} employs higher-resolution models of galaxies, 
%with a mass resolution of $500 M_{\odot}$, 
and incorporates an AR-chain integrator \citep{Mikkola1999MNRAS.310..745M,Rantala2017ApJ...840...53R,Rantala2020_mstar} and finds that only MBH seeds embedded within extended stellar systems (which remains largely unstripped during the merge)  can effectively migrate BHs to the center of the merged galaxy, where they may form a binary. In contrast, significant tidal stripping during the initial infall of the MBH host galaxy lead to naked MBH seeds in a merger remnant. In this case, MBHs stall at distances of 0.1 - 1 kpc \citep{Zhou2024MagicsII}.Therefore, stellar structures, such as NSCs surrounding the MBHs, are crucial for creating conditions that facilitate MBH seed mergers. However, the resolution limits of MAGICS-II preclude the proper examination of realistic nuclear structures around MBHs, such as NSCs, highlighting the need for high-resolution N-body simulations to accurately capture the detailed dynamics of MBHs within dense stellar environments.

 There is a wealth of literature available on modeling stellar clusters in cosmological simulations. While this is hard to do self-consistently, studies such as E-MOSAICS \citep[e.g.,][]{EMosaics2018MNRAS.475.4309P} and FIRE \citep[e.g.,][]{Fire2023MNRAS.521..124R} have employed multi-scale techniques combining high resolution simulations with cosmological simulations to study the evolution of globular clusters in galaxies. In the same vein as these previous works, we employ high resolution N-body models informed from cosmological simulations in this work.
 %Previous studies have relied on idealized models of NSCs wherein the extended galactic environment is largely ignored \tiziana{This is not quite true: you need to refer to the work of Carl R. and others too - there have been big efforts to try and simulate stellar clusters in  cosmological environment (e.g. E-MOSAIC). It is true this cannot be done fully self-consistently - but in some sort of multi-scale simulations etc.. it has been done}. 
 Upon the merger of two galaxies, the NSCs present in either galaxy would be subject to the tidal effects from the stellar bulge or the  DM halo as they sink to the center, resulting in mass loss over time and thereby affecting their sinking efficiency. A systematic study of the effect of MBH seed mergers due to NSCs in realistic galactic environments is missing. Our work, therefore, is primarily motivated to understand the efficiency of MBH seed mergers embedded in NSCs in galactic environments informed by cosmological and galaxy merger simulations. 
 
 In this work, we introduce the MAGICS-NSC suite of high-resolution $N$-body simulations, using up to $10^7$ particles, to study the evolution of MBH seeds from a separation of hundreds of parsecs to GW emission stage under the influence of NSCs. Our models are directly informed from the MAGICS-I \citep[][hereafter \citetalias{Chen2024}]{Chen2024} and MAGICS-II \citep[][hereafter \citetalias{Zhou2024MagicsII}]{Zhou2024MagicsII} suite of galaxy merger resimulations . We use the Fast Multipole Method (FMM) \citep{greengard1987fast,cheng1999fast} based $N$-body code {\tt\string Taichi} \citep{Zhu2021NewA...8501481Z,Mukherjee2021ApJ...916....9M,Mukherjee2023MNRAS.518.4801M} which has been shown to be as accurate as direct-summation based $N$-body codes while using a fraction of the computational time. Our simulations are state-of-the-art and present an effort where the evolution of the seeds is traced accurately from kiloparsec to milliparsec scales, including not only the stellar bulge but the surrounding DM halo. %\nianyi{I would be a little bit cautious in stating that this is the "first effort". If you want to say this then we should add some delimiters like "directly informed by cosmological ICs"}.

We begin by introducing the \citetalias{Chen2024} and \citetalias{Zhou2024MagicsII} resimulation suites in section \ref{sec:resim} and then briefly describe the computational methods in section \ref{sec:methods} and our models in section \ref{sec:models}. This is followed by the results in section \ref{sec:results}. We then describe some of the implications of this work in section \ref{sec:discussion} and conclude in section \ref{sec:conclusions}.
\section{The MAGICS Project} \label{sec:resim}
%\begin{figure*}
%    \begin{center}
%    \includegraphics[width=1.0\textwidth]{m6-sys1-evol.png}
%    \caption{\dipto{Figure to connect with MAGICS I to demonstrate what the resimulations look like}}
%    \label{fig:astrid_resim}
%     \end{center}
%\end{figure*}

The simulations in this work are based on the initial conditions of galaxy mergers in \citetalias{Chen2024} and \citetalias{Zhou2024MagicsII}, which are directly informed by the MBH mergers in the \texttt{ASTRID} cosmological simulation.
Here we briefly describe \texttt{ASTRID} and, \citetalias{Chen2024} and \citetalias{Zhou2024MagicsII} simulation suites.

\texttt{ASTRID} is a cosmological hydrodynamical simulation with $250 \hmpc$ per side and $2\times5500^3$ initial tracer particles comprising dark matter and baryons \citep{Bird-Astrid, Ni-Astrid}. 
The simulation includes a full-physics, sub-grid models for galaxy formation, SMBHs and their associated supernova and AGN feedback, as well as inhomogeneous hydrogen and helium reionization. 
BHs are seeded in haloes with $M_{\rm halo} > 5 \times 10^9 \hmsun$ and $M_{\rm *} > 2 \times 10^6 \hmsun$, with seed masses stochastically drawn from a power-law distribution within the mass range between $3\times10^{4} \hmsun$ and $3\times10^{5} \hmsun$, motivated by the direct collapse scenario proposed in \cite[e.g.;][]{LodatoPN2007}. 
The gas accretion rate onto the black hole is estimated via a Bondi-Hoyle-Lyttleton-like prescription \citep{DSH2005}. 
The black hole radiates with a bolometric luminosity $L_{\rm bol}$ proportional to the accretion rate $\dot{M}_\bullet$, with a mass-to-energy conversion efficiency $\eta=0.1$ in an accretion disk according to \cite{Shakura1973}. 
5\% of the radiated energy is coupled to the surrounding gas as the AGN feedback.  
Dynamics of the black holes are modeled with a sub-grid dynamical friction model \citep{Tremmel2015,Chen2021}, yielding well-defined black hole trajectories and velocities. 
We boost the dynamical mass of the black holes to $1.2\times 10^7\,M_\odot$ for gravitational interactions to alleviate the artificial dynamical heating from other particle species.
Two black holes can merge if their separation is within two times the gravitational softening length $2\epsilon_g=3\,{\rm ckpc/h}$ and they are gravitationally bound to the local potential.

\astrid has the largest MBH merger population at high-redshift \citep{Chen2022MNRAS.510..531C} with MBH masses in the range $5\times 10^{4}\,M_\odot < M_{\rm BH}<5\times 10^{10}\,M_\odot$. 
In \citetalias{Chen2024}, we select 15 out of 2107 $z\sim 6$ MBH mergers from \astrid to perform high-resolution resimulations.
The resimulations have a dark matter and gas particle mass of $8000\,M_\odot$ and a stellar particle mass of $2000\,M_\odot$.
The gravitational softening is $80\,{\rm pc}$ for dark matter and gas, $20\, {\rm pc}$ for stars, and $10\,{\rm pc}$ for black holes (the inter-species softening length is the maximum value of the two species).
The subgrid physics models in \citetalias{Chen2024} are similar to those of \astrid, except that we alleviate the boost in the MBH dynamical mass due to the much higher particle resolution.
The dynamics of MBH pairs are also followed to smaller separations ($2\epsilon_{\rm g, BH}=20\,{\rm pc}$) compared with \astrid.

To trace MBH binary dynamics down to smaller scales, 
\citetalias{Zhou2024MagicsII} uses the \texttt{KETJU} code \citep[][]{Rantala2017ApJ...840...53R,Mannerkoski2023MNRAS.524.4062M} to study 6 merging systems identified in \citetalias{Chen2024}. 
\texttt{KETJU} replaces the leapfrog integration with the algorithmically regularized \texttt{MSTAR} integrator \citep{Rantala2020_mstar} in regions around each MBH. In these regions, interactions of BH-BH, BH-star, and BH-DM are unsoftened. 
Post-Newtonian (PN) corrections up to the order of 3.5 \citep{Mora2004_PN} are included to account for general relativistic effects on the MBH binary.
In principle, this could resolve MBH evolution down to separations of tens of Schwarzschild radii. Particle splitting is implemented in \citetalias{Zhou2024MagicsII}. All particles, including DM, gas and stars, within $1$ kpc from the binary center of mass are split to $500\ \mathrm{M}_{\odot}$. 
The gravitational softening is 20 pc for gas particles, and 5 pc for DM, stellar, and BH particles.

\section{Computational Methods} \label{sec:methods}
To simulate large-$N$ systems within a reasonable amount of time,
we utilize the FMM \citep[e.g.,][]{greengard1987fast,cheng1999fast} based code {\tt\string Taichi} \citep{Zhu2021NewA...8501481Z, Mukherjee2021ApJ...916....9M, Mukherjee2023MNRAS.518.4801M}. \cite{Mukherjee2021ApJ...916....9M} showed that {\tt\string Taichi} can simulate systems as accurately as direct-summation based collisional $N$-body codes while scaling as $\mathcal O(N)$. The accuracy of the force calculation in {\tt\string Taichi} can be tuned via the usage of an input accuracy parameter ($\epsilon$) which controls the opening angle, and a multipole expansion parameter ($p$) which controls the number of expansion terms used in the force calculation. Using {\tt\string Taichi} we can simulate large-$N$ systems without the usage of specialized hardware within a physically reasonable amount of time. \citet{Mukherjee2023MNRAS.518.4801M} extended {\tt\string Taichi} to include a fourth-order force-gradient integrator using the {\tt\string HHS-FSI} scheme \citep{Rantala2021MNRAS.502.5546R} and regularization using the {\tt\string AR-Chain-Sym6+} scheme \citep{Mikkola1999MNRAS.310..745M, Wang2021MNRAS.505.1053W} using the {\tt\string SpaceHub} library. {\tt\string Taichi} includes adaptive, individual time-symmetrized timesteps which help conserve energy better than non time-symmetrized schemes \citep{Pelupessy2012NewA...17..711P}. 

When the separation between the MBHs, $\Delta r$, is $> 30$ pc, we set $\epsilon = 2\times 10^{-5}$, $p=12$, and a timestep parameter $\eta = 0.3$. In this stage, we do not use regularization to save on computational expenses. Our experiments find that regularization does not make a significant difference in this era. For smaller separations with $\Delta r\leq 30$ pc, we set $\epsilon = 2\times 10^{-6}$, $p=15$, and a timestep parameter $\eta \leq 0.1$ and enable regularization. This ensures an accurate evolution of the MBH binary in the three-body hardening stage.

The interactions of the MBHs with other particles are never softened. However, we soften the interactions between particles other than the BHs using a Plummer-type softening. Different particle types have different softening lengths. The softening length for interactions between particle types $i$ and $j$, $\varepsilon_{ij}$, is determined as:

\begin{equation}
    \varepsilon_{ij} = \sqrt{\frac{{\varepsilon_i}^2 + {\varepsilon_j}^2}{2}} 
\end{equation}
where $\varepsilon_i$ is the softening length of particle type $i$ and $\varepsilon_j$ is the softening length of particle type $j$. 
We do extensive experimentation to understand the effect of softening and find that there are minimal differences in the orbit of the MBHs in the simulations that use softening versus those that do not. We re-emphasize that the interactions of MBHs with other particles are \textit{never} softened in either case. The values of softening used for different particle types are provided in table \ref{tab:softening} in section \ref{sec:models}.

 Despite the average simulation containing $4 \mathrm{m} - 8 \mathrm{m}$ particles, our simulations take about 7-14 days of wall-clock time to run, which includes the evolution in the hard binary stage. All simulations presented in this study were performed using 48-64 threads on an {\tt\string AMD Epyc 7742} node. Energy in all of our simulations is conserved to the order of 0.01\% - 0.1\%. Extra care is taken to ensure that the energy error is always at least an order of magnitude below the fraction of hard binary's energy to the total energy of the system.
\section{Models} \label{sec:models}
\begin{figure}
    \begin{center}
    \includegraphics[width=0.5\textwidth]{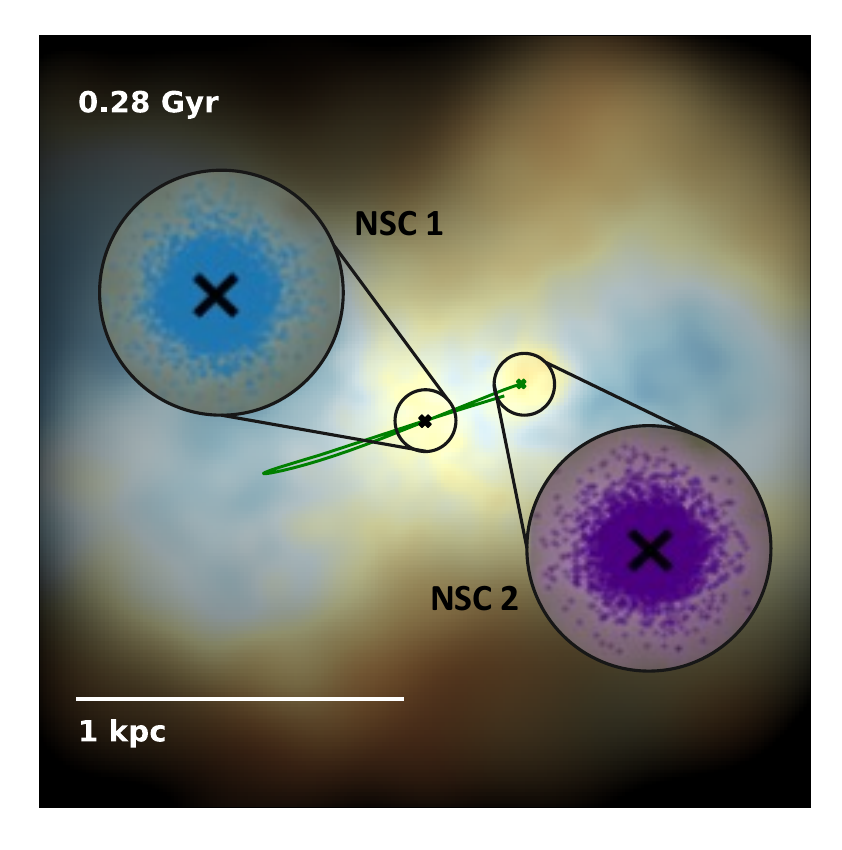}
    \caption{Visualization of the stellar density field of system 12 from the \citetalias{Zhou2024MagicsII} suite. Brightness represents density, while color indicates stellar age: blue for younger stars and yellow for older stars. Overdense regions are observed around the MBHs (crosses), marking the nuclei of the original galaxies. These nuclei correspond to nuclear clusters surrounding the MBHs, but they are less dense and less massive than realistic NSCs. To assess the impact of denser and more massive NSCs, we introduce NSCs (blue and purple particles in the zoomed-in circles) around the two MBHs and continue their evolution in this study. }
    \label{fig:clusters_in_magics2}
     \end{center}
\end{figure}

Our initial conditions are derived directly from the \citetalias{Chen2024} and \citetalias{Zhou2024MagicsII} suites of galaxy merger resimulations. The positions and velocities of all particles are extracted from the \citetalias{Chen2024} $N$-body data. Once the data is set up and the NSCs are added, our simulations proceed in two stages: \texttt{STAGE-I} where the system is evolved until the MBHs reach a separation of $\Delta r \approx 30$ pc, and \texttt{STAGE-II} where the MBHs subsequently form a hard binary and shrink further via three-body interactions. To determine the hard-binary radius $r_{\rm h}$, we first find the influence radius $r_{\rm infl}$. Following \citet{Merritt2013degn.book.....M} equation (8.71), the influence radius of the binary $r_{\rm infl}$ is defined as  
\begin{equation} \label{equation:influence_rad_binary}
    r_{\rm infl} \equiv r_{\rm enc}( 2 (M_{p}+M_{s}))
\end{equation}
where $M_p$ and $M_s$ are the masses of the primary and secondary MBHs respectively, and $r_{\rm enc}$ is the radius enclosing twice the mass of the binary. We note that $M_1$ and $M_2$ are also used to denote the masses of the MBHs in the simulations. When the MBHs get sufficiently close to begin the process of formation of a bound binary, we switch to $M_s$ and $M_p$. The hard binary radius $r_{\rm h}$ can then be determined as 
\begin{equation} \label{equation:hard_binary_binary}
    r_{\rm h} = \frac{q}{(1+q)^2} \frac{r_{\rm infl}}{4}.
\end{equation}
where $q \equiv \frac{M_s}{M_p}$ is the mass-ratio of the MBH binary.

Dynamical processes such as DF and three-body hardening are sensitive to the MBH mass to particle mass ratio \citep[e.g.,][]{Pfister2019MNRAS.486..101P,Genina2024arXiv240508870G}. To ensure that the ratio is large enough, we perform particle splitting in the same manner as \citet{Khan2012ApJ...756...30K,Khan2016ApJ...828...73K}. We then add the NSCs around the MBHs, assuming that they follow the \citet{Dehnen1993MNRAS.265..250D} density profile with a shallow $\gamma=0.5$ cusp. We describe the steps used to generate the initial conditions in further details below.

\subsection{System and snapshot selection}

We select a subset of 6 systems from the original \citetalias{Chen2024} set where the MBH binaries merge within 1.2 Gyr. We also simulate an additional system, system 6, from the \citetalias{Chen2024} set in order to expand the range of environments in which the mergers happen. Once the systems are chosen, we identify the time when the separation between the MBHs is greater than 300 pc and the gas density is about an order of magnitude lower than the stellar and DM density within a kiloparsec from the potential center of the galaxy. %\yihao{the latter part of this sentence ``the gas density is...'' is a little confusing to me and maybe you can clarify it a bit more. Is there any evolution of the gas density profile? Is there a timepoint when the gas density profile switches from higher than the stellar profile to lower? Or do you mean you make sure at this point, the gas density profile is lower than the stellar/DM density profile?}
This is done since {\tt\string Taichi} cannot handle gas effects. The lower gas density ensures that active star formation in the inner few hundred parsecs does not significantly affect our results. 

\subsection{Particle splitting}

The stellar, gas, and DM particle masses in the resimulations suite are $2\times10^3 M_{\odot}, 8\times10^3 M_{\odot}, \mathrm{ and } \,8\times10^3 M_{\odot}$ respectively. This is somewhat insufficient to resolve DF and three-body hardening effects directly. To increase the particle mass resolution, we split each particle into more particles. Particle splitting is performed in the same manner as \citet{Khan2012ApJ...756...30K,Khan2016ApJ...828...73K}. Each DM or gas particle is successively split over a uniform sphere of radius equaling its softening length until the mass of each split particle is $500 M_{\odot} - 1000 M_{\odot}$ while each stellar particle is split until the mass of each split particle is $250 M_{\odot} - 500 M_{\odot}$. The split particles are assigned the same velocity as that of the parent particle. This conserves energy and momentum. In each system, the mass ratio of the least massive MBH to the particle mass is ensured to be greater than 100. 

Splitting increases the number of particles in the simulations by factors of $5-20$. To ensure that the simulations can be performed within a physically reasonable amount of time, splitting is only performed within $1$ kpc from the potential center of the galaxy. Afterwards, radial cut at a distance of $2$ kpc is performed. Experiments are performed to ensure that the radial cuts did not significantly affect the dynamics of the binary or the overall mass profile in the region of interest. The above procedures result in each system containing $N \approx 4\times10^6 - 8\times10^6$. The systems are evolved until the MBHs reach $\Delta r = 30$ pc. This marks the end of \texttt{STAGE-I} of our simulations. 

When the MBHs reach $\Delta r = 30$ pc, \texttt{STAGE-II} begins. We perform a radial truncation of the particle dataset at 1 kpc from the minimum potential center and perform particle splitting again to ensure that \textit{all} particles are $250 M_{\odot} - 500 M_{\odot}$. In our higher resolution models, the stellar particles are split until they have a mass of $62.5 M_{\odot}$. The three-body evolution is sensitive to the mass-ratio of the particle to the secondary, necessitating the particle splitting procedure for all particles. This results in a sufficient resolution to resolve the binary binding stage and the three-body interactions that lead to hardening of the binary over time. In most models, this leads to $N_{\mathrm{hb}} \approx 3.5\times10^6 - 7.5 \times 10^6$, where $N_{\mathrm{hb}}$ is the number of particles in the hard binary stage. Across our suite of simulations, we are able to achieve $ 10^2 < M_{\rm MBH} / M_{*} < 1.5 \times 10^3$  The mass ratio of the secondary MBH to that of the stellar particles is comparable to the values provided in \citet{Khan2021MNRAS.508.1174K}. For most models, the evolution of the MBHs is followed down to a semi-major axis $a \leq r_{\rm h} / 10$.

The split particles are assigned softening values as denoted in table \ref{tab:softening} whereas the unsplit particles in \texttt{STAGE-I} of the simulations retain the softening values from the original \citetalias{Chen2024} set. The NSC stars are assigned zero softening.

% \usepackage{graphicx}
\begin{table}
\centering
\resizebox{0.45\textwidth}{!}{%
\begin{tabular}{ccc} \hline
Simulation stage & Particle type & Softening length \\ \hline
\texttt{STAGE-I}  & \begin{tabular}[c]{@{}l@{}}Bulge stars\\ NSC stars\\ Gas\\ DM\end{tabular} & \begin{tabular}[c]{@{}l@{}}5 pc\\ 0 pc\\ 5 pc\\ 20 pc\end{tabular}       \\ \hline
\texttt{STAGE-II}         & \begin{tabular}[c]{@{}l@{}}Bulge stars\\ NSC stars\\ Gas\\ DM\end{tabular} & \begin{tabular}[c]{@{}l@{}}0.01 pc\\ 0 pc\\ 0.01 pc\\ 10 pc  \end{tabular}  \\ \hline
\end{tabular}%
}
\caption{List of inter-particle softening values used for different particle types that undergo splitting in our simulations. Plummer-type softening is incorporated into the force calculations. The interactions between the NSC stars are never softened. In \texttt{STAGE-II} we reduce the softening to better resolve loss-cone scattering. We remind the reader that the interactions between MBHs and other particles are never softened. }
\label{tab:softening}
\end{table}

\subsection{Generating the nuclear star clusters}
\begin{table*}
\centering
\resizebox{\textwidth}{!}{%
\begin{tabular}{ccccccccc}
\hline
MAGICS-I/II system name & Model name  & $N_{\rm gal}$ [$10^6$] & $N_{\rm NSC}$ [$10^3$] & $M_{\mathrm{*,gal}}$ [$10^8 M_{\odot}$]  &$M_{\mathrm{NSC,1}}$ [$ 10^6 M_{\odot}$] & $M_{\mathrm{NSC,2}}$ [$10^6 M_{\odot}$] & $M_{\mathrm{MBH,1}}$ [$10^4 M_{\odot}$] & $M_{\mathrm{MBH,2}}$ [$10^4 M_{\odot}$] \\ \hline
System 1  & \hspace{-10pt} \begin{tabular}[c]{@{}l@{}}\texttt{sys1\_FRe\_O}\end{tabular} &  7.7   &   \begin{tabular}[c]{@{}l@{}}44\end{tabular}    &   2.7    &  \begin{tabular}[c]{@{}l@{}}9.03\end{tabular}    &   \begin{tabular}[c]{@{}l@{}}1.88\end{tabular}   &  6.00    &    16.7      \\ \hline
System 2  & \hspace{-10pt} \begin{tabular}[c]{@{}l@{}}\texttt{sys2\_FRe\_O}\\ \texttt{sys2\_VRe\_O}\end{tabular} & 6.2  &   \begin{tabular}[c]{@{}l@{}}97\end{tabular}   &   9.2    &  \begin{tabular}[c]{@{}l@{}}41.1\end{tabular}    &  \begin{tabular}[c]{@{}l@{}}7.20\end{tabular}    &  13.2    &    26.7      \\ \hline
System 3  & \hspace{-10pt} \begin{tabular}[c]{@{}l@{}}\texttt{sys3\_FRe\_O}\\ \texttt{sys3\_FRe\_HI}\end{tabular} & 5.2  &  \begin{tabular}[c]{@{}l@{}}10\\ 80\end{tabular}     &  1.2     & \begin{tabular}[c]{@{}l@{}}2.45\end{tabular}     &   \begin{tabular}[c]{@{}l@{}}2.26\end{tabular}   &   8.69   &    5.15      \\ \hline
System 6  & \hspace{-10pt} \begin{tabular}[c]{@{}l@{}}\texttt{sys6\_FRe\_O} \end{tabular} & 3.0  &  \begin{tabular}[c]{@{}l@{}}8\end{tabular}     &   1.9    &   \begin{tabular}[c]{@{}l@{}}2.34\end{tabular}   &   \begin{tabular}[c]{@{}l@{}}1.26\end{tabular}   &  25.4    &    12.5      \\ \hline
System 7  & \hspace{-10pt} \begin{tabular}[c]{@{}l@{}}\texttt{sys7\_FRe\_O} \end{tabular} & 5.6  &  \begin{tabular}[c]{@{}l@{}}17\end{tabular}     &   1.8    &  \begin{tabular}[c]{@{}l@{}}5.16\end{tabular}   &  \begin{tabular}[c]{@{}l@{}}3.25\end{tabular}    &  14.5    &    8.65      \\ \hline
System 10 & \hspace{-10pt} \begin{tabular}[c]{@{}l@{}}\texttt{sys10\_FRe\_O}\\ \texttt{sys10\_VRe\_O}\end{tabular} & 5.7   &  \begin{tabular}[c]{@{}l@{}}34\end{tabular}     &   6.8    &    \begin{tabular}[c]{@{}l@{}}12.3\end{tabular}  &  \begin{tabular}[c]{@{}l@{}}4.67\end{tabular}    &   19.5   &    25.9      \\ \hline
System 12 & \hspace{-10pt} \begin{tabular}[c]{@{}l@{}}\texttt{sys12\_FRe\_O}\\ \texttt{sys12\_FRe\_HI} \\ \texttt{sys12\_LOWM}\end{tabular} & 5.8  &  \begin{tabular}[c]{@{}l@{}}11\\ 88 \\ 11 \end{tabular}     &    1.8   &  \begin{tabular}[c]{@{}l@{}}2.65 \\ 2.65 \\ 0.35\end{tabular}    &   \begin{tabular}[c]{@{}l@{}}2.65 \\ 2.65 \\ 0.35\end{tabular}   &  54.3    &    8.76      \\ \hline
\end{tabular}%
}
\caption{Initial conditions for \texttt{STAGE-I} of our simulations. The columns denote the number of particles in the galaxy $N_{\rm gal}$ after splitting and truncation including stellar, DM, and gas particles, the total number of particles used to model both NSCs $N_{\rm NSC}$, the stellar mass of the galaxy $M_{\rm *,gal}$, masses of each NSC $M_{\rm NSC,1}$ and $M_{\rm NSC,2}$ and the MBHs present in those NSCs. $N_{\rm gal}$ for a system is kept constant between different models for each system. $N_{\rm NSC}$ depends on the mass resolution of each NSC particle. For the original resolution \texttt{O} models, we set the particle mass to be $500 M_{\odot}$ ($250 M_{\odot}$ for system 1) whereas in the higher resolution \texttt{HI} and low mass \texttt{LOWM} model we set that to $62.5 M_{\odot}$. Our average simulations consists of $N \sim 6\times10^6$ particles in this stage.   }
\label{tab:stage_i_ic}
\end{table*}

\begin{figure*}
    \begin{center}
    \includegraphics[width=0.9\textwidth]{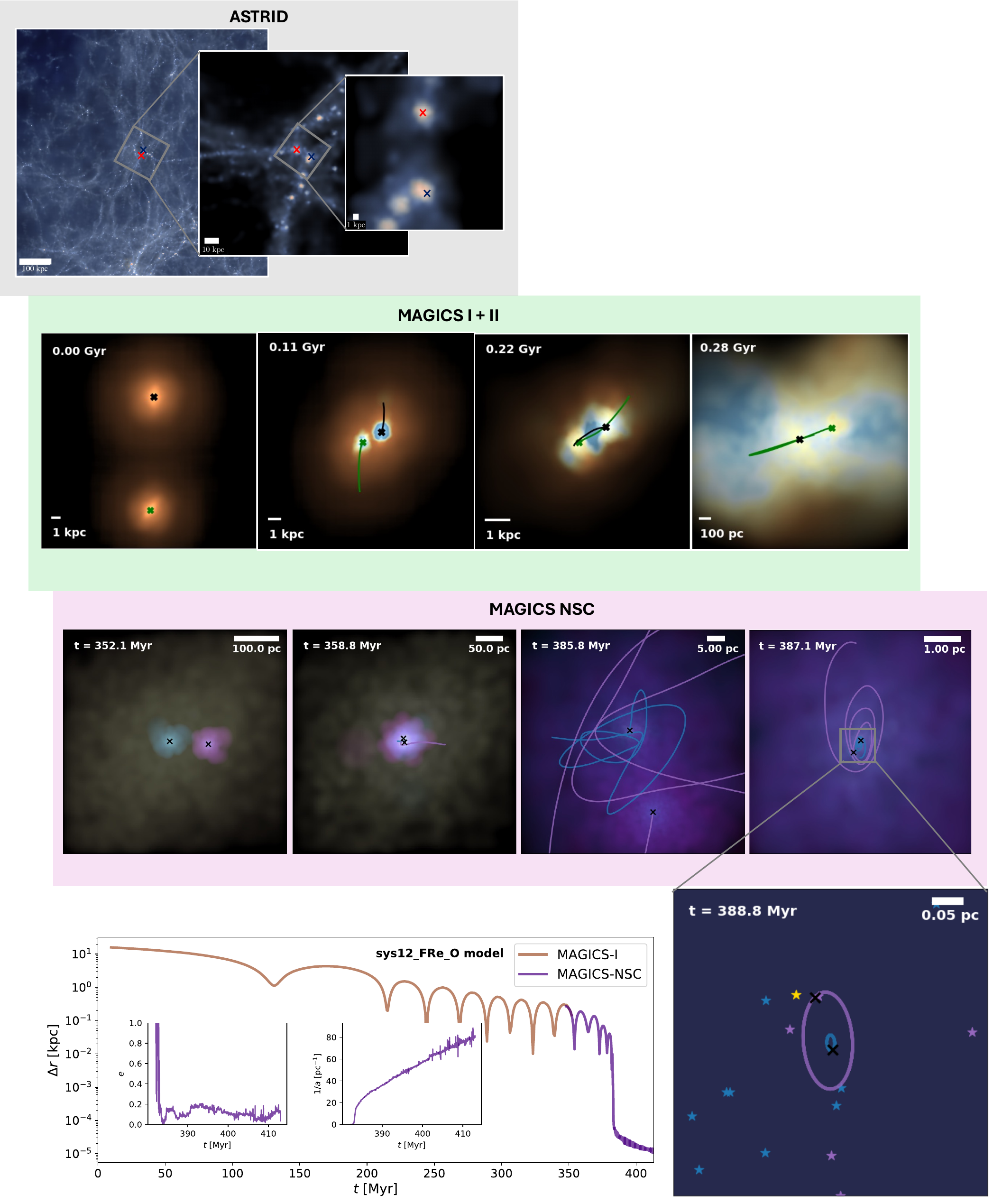}
    \caption{Visualization of the MBHs from kpc to mpc scales for system 12. Top: Zoom-in of the \texttt{ASTRID} volume showing the galaxies hosting MBHs (crosses), as resimulated in 
 \citetalias{Chen2024} and \citetalias{Zhou2024MagicsII}. Middle: High-resolution merger resimulation from \citetalias{Zhou2024MagicsII} displaying two galaxies from \texttt{ASTRID} with brightness indicating density and color indicating age (blue for younger stars, red for older). MBHs reach a separation of $\sim 300$ pc within$\sim 300$Myr.
 Bottom: Central 600 pc of the galaxy, embedding two seeds inside NSCs (blue and purple spheres) within the stellar bulge background (yellow). Interactions between NSCs enhance MBH sinking, forming a hard binary by 388.8 Myr.
Bottom left: Evolution of the binary separation $\Delta r$ as a function of the time $t$. The initial phase (brown line) is followed in \citetalias{Chen2024} whereas the later phase with the MBHs inside clusters (violet line) is followed in this work. Insets show the evolution of the eccentricity $e$ and inverse semi-major axis $1/a$ after a bound binary has formed. Bottom right: Binary visualization at a few mpc, showing the secondary's orbit (purple line) around the primary (blue line). Star colors indicate origins: blue (left NSC), purple (right NSC), yellow (bulge).
}
    \label{fig:nsc_system_evol}
     \end{center}
\end{figure*}

In \citetalias{Zhou2024MagicsII}, the authors find that the presence of extended stellar systems surrounding the MBH seeds are critical to the sinking process. These stellar systems represent the unstripped nuclei of the original galaxies that grow over time due to star formation. A visualization of these stellar systems for system 12 from their simulations is provided in Figure \ref{fig:clusters_in_magics2}. The authors find that when the extended stellar systems surrounding the MBH seeds are fully stripped, the MBHs stall at large separations and are unable to efficiently sink to the center to form a binary. 

 The finite particle limit and force softening hinder the accurate resolution of nuclear substructures such as nuclear star clusters (NSCs) and their dynamical effects in \citetalias{Zhou2024MagicsII}. The limited mass resolution introduces mass loss, primarily due to artificial stripping caused by two-body interactions. This energy loss is proportional to $\frac{\ln(N)}{N}$, where $N$ is the number of particles. Consequently, achieving a more accurate estimation of stellar system mass loss over time necessitates higher particle numbers. Additionally, the softening length employed for stellar particles results in a cored density profile within $4 \times \epsilon_s$ ($\sim$ 100 pc), where $\epsilon_s$ denotes the softening length. Studies have shown that such cored density profiles are subject to enhanced tidal stripping \citep[e.g.,][]{Errani2021MNRAS.505...18E,Du2024PhRvD.110b3019D}, leading to a significant reduction in density after a few orbits. In contrast, in the presence of an NSC, we expect a cuspy profile in the central few parsecs of the nucleus. These cuspy profiles are more resistant to tidal stripping and tend to retain a significant fraction of their mass, suggesting that, under ideal conditions, the NSC would survive. The aforementioned limitations lead to the presence of an extended stellar system only surrounding system 12. A detailed analysis of the stellar profile of system 12 shows that the mass ratio of the two galactic nuclei is initially equal. In contrast, other systems with unequal mass ratios experience complete disruption of the less massive nucleus during the merger process, leading to the formation of a naked MBH seed. However, this outcome is an artifact of force softening, as the presence of initial cusps, representative of NSC, would have resulted in their survival. This observation motivates further investigation into the effects of NSCs in systems that experienced stripped nuclei in this work.
%However, the nuclei surrounding the MBHs are underdense and less massive compared to observed NSCs which motivates us to study how denser and more massive NSCs affect the inspiral.

As a first step, we analyze system 12 from \citetalias{Zhou2024MagicsII} to determine the mass contained within 50 pc of each MBH when they are more than 100 pc apart. We find that roughly $3.5\times10^5 M_{\odot}$ is contained within each nucleus. We, then, assume the mass represents an NSC and follows the \citet{Dehnen1993MNRAS.265..250D} density profile which is defined as
\begin{equation}
    \rho(r) = \rho_0 \left( \frac{r}{r_{0}} \right)^{-\gamma_{\rm cl}} \left( 1+\frac{r}{r_{0}} \right)^{\gamma_{\rm cl}-4}
\end{equation}
where $\rho_0$ is a normalization parameter, $r_0$ is a scale radius, and $\gamma_{\rm cl}$ controls the slope of the inner density profile.
We set $\gamma_{\rm l}=0.5$ indicating a shallow inner cusp, and $r_0 = 1.4$ pc. $r_0$ is related to the effective radius $R_{\rm eff}$ of the NSCs as
\begin{equation}
    R_{\rm eff} \approx 0.75 r_0 \left ( 2^{1/(3-\gamma_{\rm cl})} - 1 \right )^{-1}.
\end{equation}
For our choice of $r_0$, we find $R_{\rm eff}=3.3$ pc.

The NSCs are generated taking into account the cluster potential and the MBH potential. We perform experiments to understand the effect of the galactic potential on the stability of our profiles and we find that our NSCs remain sufficiently stable as the galactic potential is subdominant compared to the cluster and MBH potentials. The NSCs are assumed to be spherical and isotropic initially and are generated using \texttt{Agama} \citep{Vasilev2019MNRAS.482.1525V}. They are then placed around the MBHs. We note that while the mass of each NSC in this case is determined from \citetalias{Zhou2024MagicsII}, the positions and velocities of other particles, including the MBHs, are determined from the same system in the \citetalias{Chen2024} set. This is done in order to be consistent with the other models. This method produces a lower bound on the mass estimates for the NSCs and this model is labeled as \texttt{sys12\_LOWM}. The mass of each NSC particle is set to $62.5 M_{\odot}$ for this particular model. Given that the effective radii of typical NSCs are comparable to, or less than the stellar softening length used in the \citetalias{Zhou2024MagicsII} simulations, the mass estimates are a lower bound on NSC masses.  

To better ascertain the masses of the NSCs used in other models, we follow a simple prescription using the particle dataset from \citetalias{Chen2024}. We ensure that our masses are informed from the galactic environment by assuming that the total initial mass of both NSCs is equal to the mass contained within 100 pc from the center of mass of the MBH binary at the end of \citetalias{Chen2024} resimulations. We compare the obtained NSC mass $M_{\rm NSC}$ to the galaxy stellar mass $M_{\rm *,gal}$ for each model to ensure that it is physically realistic. We find  that the total NSC mass is on the upper end of the $M_{\rm{*,gal}}-M_{\rm{NSC}}$ mass relation from \citet{Neumayer2020A&ARv..28....4N} and is also consistent with the $M_{*,\rm{gal}} - M_{\rm NSC}$ relationship from \citet{Pechetti2020ApJ...900...32P}. 

Once the total mass in NSCs for each system is found, we then determine the mass of each individual NSC in a particular system. The masses of the two NSCs, $M_{\rm NSC,1}$ and $M_{\rm NSC,2}$, are based on the mass ratio of the inner 100 pc of the stellar bulge of each galaxy right before the galaxies merge. This helps capture some information related to initial star formation in the centers of each galaxy. Once the mass of each individual NSC has been found, we follow the same steps as before to generate $N$-body representations assuming a \citet{Dehnen1993MNRAS.265..250D} density profile with a $\gamma=0.5$ cusp. 
For simplicity  we first generate models where we set $r_0 = 1.4$ pc for all clusters, thereby fixing the effective radius. We label these models as \texttt{FRe} models indicating all such NSCs have an initial \textit{fixed} $R_{\rm eff}$. Although $R_{\rm eff}$ stays constant as a function of $M_{\rm NSC}$, the generated clusters are consistent with observed NSCs with similar masses \citep[e.g.,][]{Georgiev2016MNRAS.457.2122G}. To understand the effect of changing $R_{\rm eff}$ as a function of NSC mass, we perform simulations on a subset of systems, systems 2 and 10, where we calculate the initial $R_{\rm eff}$ of the clusters using the following relation from \citet{Pechetti2020ApJ...900...32P}:

\begin{equation} \label{equation:pechetti_reff}
    \rm{log}(R_{\rm eff}) = 0.33 \rm{log} \left ( \frac{M_{\rm NSC}}{10^6 M_{\odot}} \right) + 0.36.   
\end{equation}
These models are labeled with \texttt{VRe} indicating that the NSCs have a \textit{varying} $R_{\rm eff}$. 
While the overall generation of the clusters is not fully self consistent since we are adding extra mass to the initial system, we find that in absence of tidal interactions between clusters, the NSCs remain quite stable. 

For most of our models, the mass of each individual NSC particle is set to be $500 M_{\odot}$. For better resolution, we set the NSC particle mass to be $250 M_{\odot}$ for system 1. These models are labeled with \texttt{O} suffix. For two of our systems, systems 3 and 12, we generate higher resolution NSCs to test the convergence of our results. The mass of each NSC particle in these models are set to be $62.5 M_{\odot}$. The higher resolution models are labeled with \texttt{HI} suffix. In all of our models, the NSC particles are set to have zero softening. The initial conditions for all our models for both stages of the simulations are summarized in tables \ref{tab:stage_i_ic} and \ref{tab:stage_ii_ic}.

% \usepackage{graphicx}

\begin{table}
\centering
\resizebox{0.4\textwidth}{!}{
\begin{tabular}{ccc} \hline
Model & $N [10^6]$ & $M_{s} / M_{*}$ \\ \hline
\begin{tabular}[c]{@{}l@{}}\texttt{sys1\_FRe\_O}\end{tabular} & \begin{tabular}[c]{@{}l@{}} 4.49\end{tabular} & \begin{tabular}[c]{@{}l@{}} 240\end{tabular} \\ \hline
\begin{tabular}[c]{@{}l@{}}\texttt{sys2\_FRe\_O} \\ \texttt{sys2\_VRe\_O}\end{tabular} & \begin{tabular}[c]{@{}l@{}} 5.16 \\ 5.46 \end{tabular} & \begin{tabular}[c]{@{}l@{}} 266 \end{tabular} \\ \hline 
\begin{tabular}[c]{@{}l@{}}\texttt{sys3\_FRe\_O} \\ \texttt{sys3\_FRe\_HI}\end{tabular} & \begin{tabular}[c]{@{}l@{}} 4.19 \\ 7.59 \end{tabular}  & \begin{tabular}[c]{@{}l@{}} 104 \\ 825\end{tabular} \\ \hline
\begin{tabular}[c]{@{}l@{}}\texttt{sys6\_FRe\_O}\end{tabular} & \begin{tabular}[c]{@{}l@{}} 4.08 \end{tabular}  & \begin{tabular}[c]{@{}l@{}} 250 \end{tabular} \\ \hline
\begin{tabular}[c]{@{}l@{}}\texttt{sys7\_FRe\_O}\end{tabular} & \begin{tabular}[c]{@{}l@{}} 3.89 \end{tabular} & \begin{tabular}[c]{@{}l@{}} 173\end{tabular} \\ \hline
\begin{tabular}[c]{@{}l@{}}\texttt{sys10\_FRe\_O} \\ \texttt{sys10\_VRe\_O}\end{tabular} & \begin{tabular}[c]{@{}l@{}} 4.19 \\ 4.15 \end{tabular} & \begin{tabular}[c]{@{}l@{}} 390\end{tabular} \\ \hline
\begin{tabular}[c]{@{}l@{}}\texttt{sys12\_FRe\_O} \\ \texttt{sys12\_FRe\_HI} \\  \texttt{sys12\_LOWM}\end{tabular} & \begin{tabular}[c]{@{}l@{}} 3.86 \\ 10.2 \\ 7.46 \end{tabular} & \begin{tabular}[c]{@{}l@{}} 175 \\ 1401 \\ 1401\end{tabular} \\ \hline
\end{tabular}
}
\caption{Initial conditions for \texttt{STAGE-II} of our simulations. Once the MBHs reach a separation of $\Delta r \leq 30$ kpc, the second stage of our simulation begins. We perform further truncation to 1 kpc and ensure \textit{all} particles have a mass of $500 M_{\odot}$ ($250 M_{\odot}$ for \texttt{sys\_1\_FRe\_O}) in the \texttt{O} models. For the {HI} models, all stellar particles are split until they have a mass of $62.5 M_{\odot}$. The mass ratio of the secondary to the stellar particles $M_s / M_*$ is always greater than 100 for a sufficiently smooth evolution in the hard-binary stage. }
\label{tab:stage_ii_ic}
\end{table}

\section{Results} \label{sec:results}
\begin{figure*}
\centering
\begin{subfigure}[b]{\linewidth}
\includegraphics[width=\linewidth]{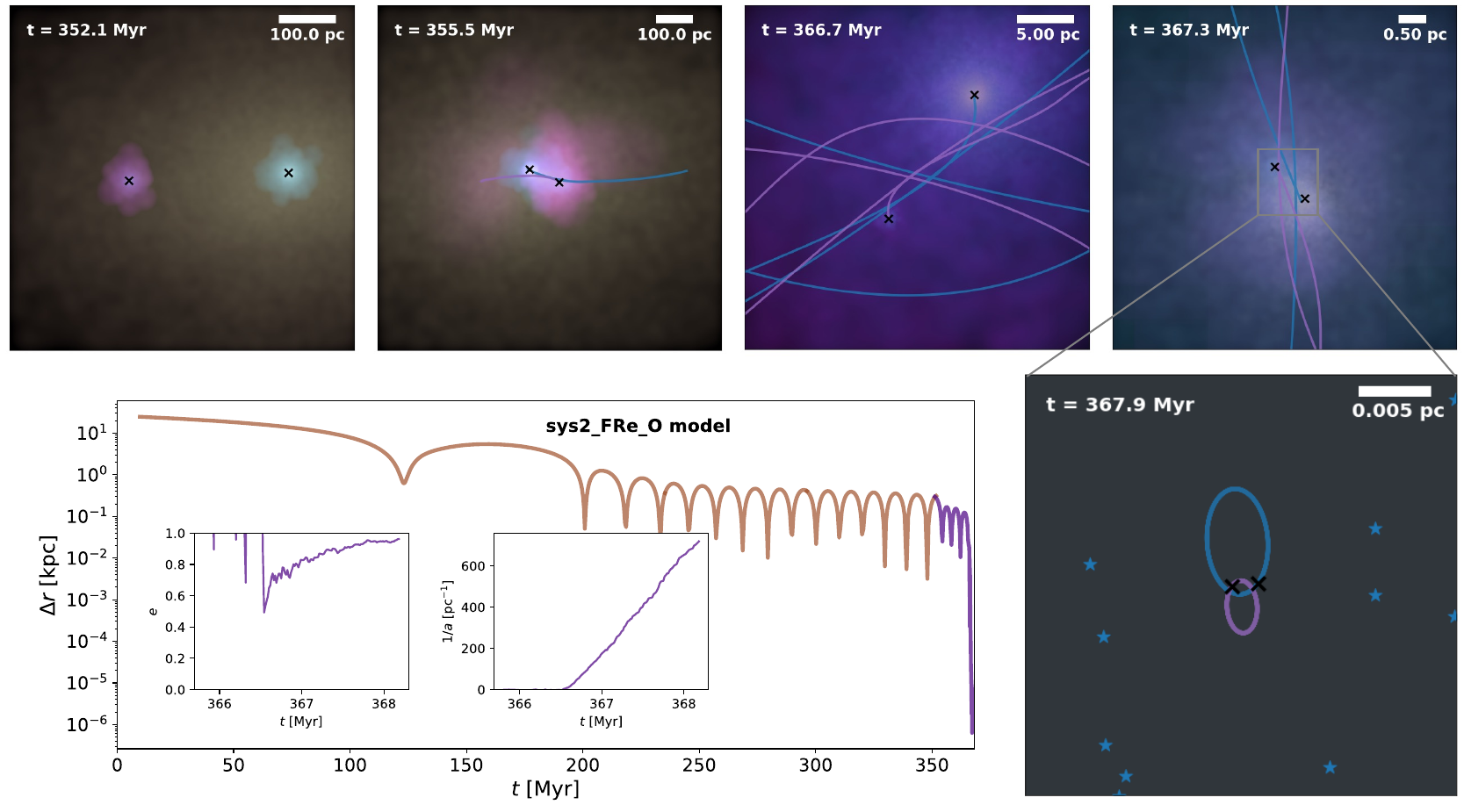}
\end{subfigure}
\begin{subfigure}[b]{\linewidth}
\includegraphics[width=\linewidth]{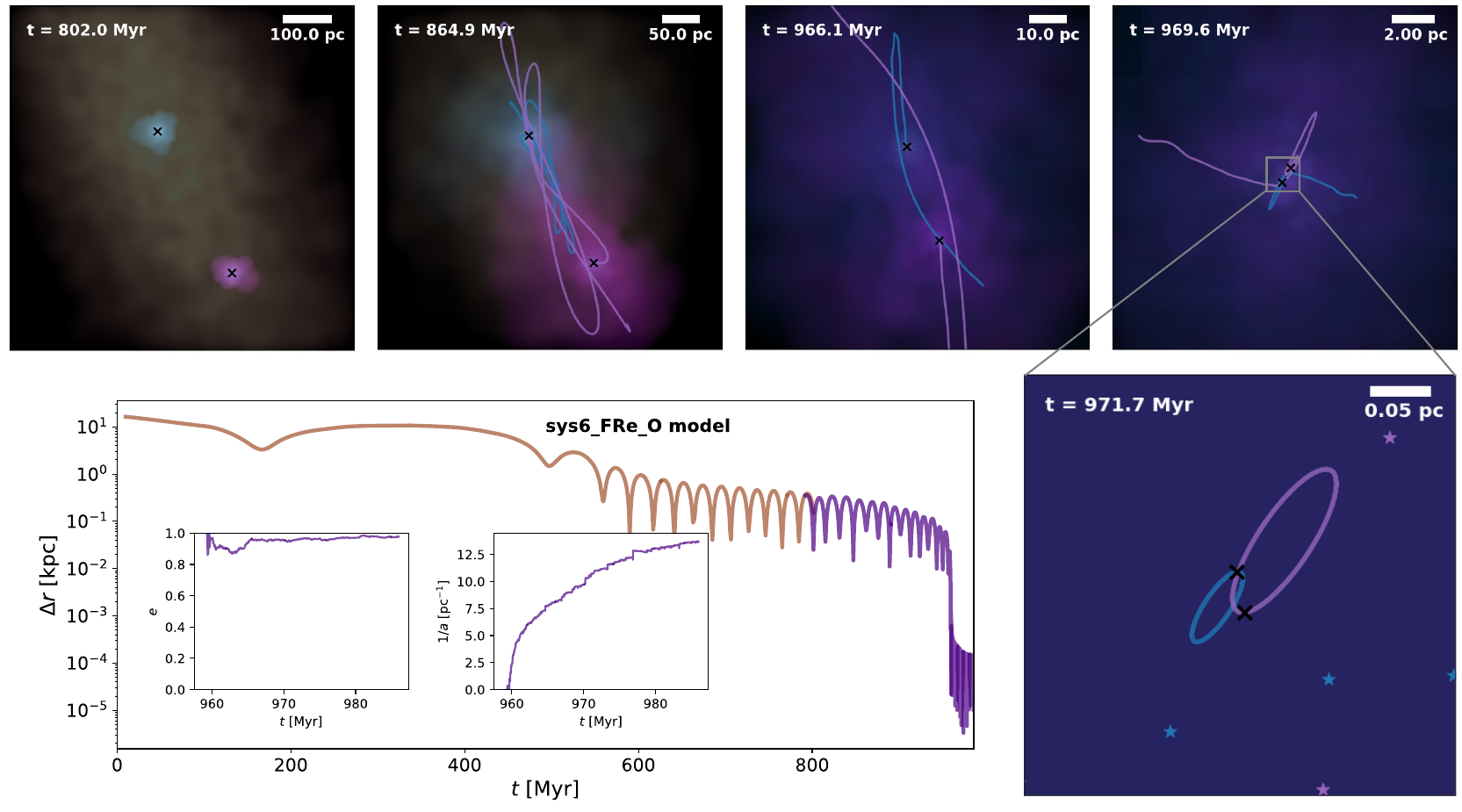}
\end{subfigure}
\caption{Visualization of two different models: \texttt{sys2\_FRe\_O} and \texttt{sys6\_FRe\_O} from our suite of simulations. The color scheme is same as that used in Figure \ref{fig:nsc_system_evol}. Top: \texttt{sys2\_FRe\_O}, a fast shrinking system. Due to the high stellar density of the surrounding galactic stellar medium and the large NSC masses, the binary shrinks quickly forming a hard binary with 15 Myr from the start of the simulation. As shown in the inset axes, the formed binary reaches a high eccentricity of 0.95 by 368 Myr while hardening to $a \sim 10^{-3}$ pc.  Bottom: \texttt{sys6\_FRe\_O}, a slow shrinking system. Contrary to the other system, the MBHs take 85 Myr to shrink to a bound binary stage in this case. The formed binary is very eccentric (0.98-0.99) as the other model (\texttt{sys2\_FRe\_O}) but the hardening rate is about three orders of magnitude lower. Consequently, the binary is only able to shrink to $\sim 100$ mpc by the end of our simulation.}
\label{fig:system_graphics}
\end{figure*}

\subsection{Initial evolution and sinking of the MBHs}

\begin{figure*}
    \begin{center}
    \includegraphics[width=1.0\textwidth]{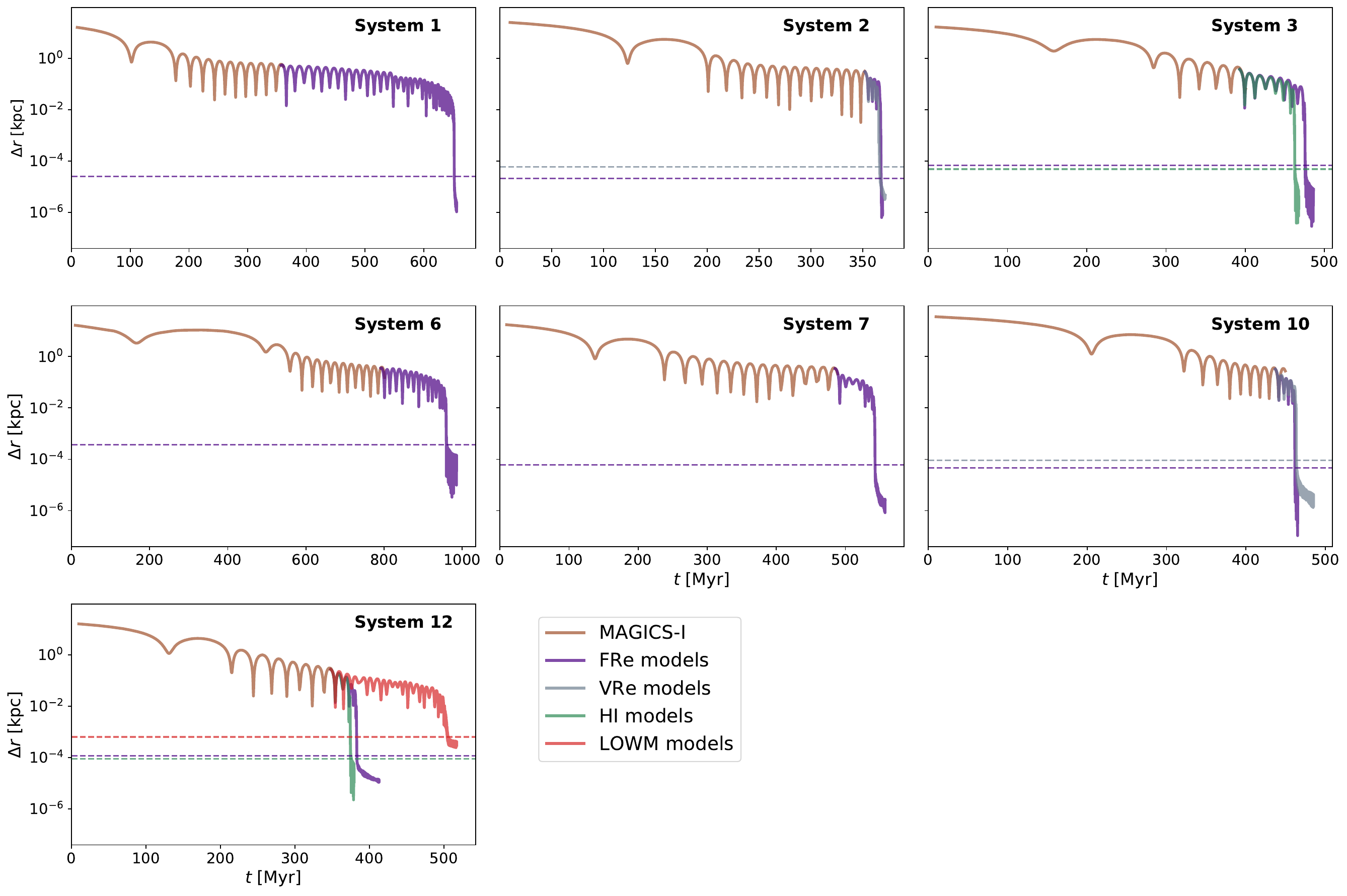}
    \caption{The evolution of the MBH separation $\Delta r$ as a function of time $t$ for all our models. Across all models, the binaries shrink to below the hard binary radius (dashed line) indicating that NSCs are efficient at accelerating MBH sinking times. The initial effective radii $R_{\rm eff}$ of the NSCs are \textit{fixed} in the \texttt{FRe} models (purple lines) whereas they are \textit{varied} in the \texttt{VRe} models (grey) following equation \ref{equation:pechetti_reff}. The \texttt{VRe} models show qualitatively similar evolution to that of the \texttt{FRe} models with differences of at most $1-2$\% in the sinking times. We notice that our higher resolution models \texttt{HI} models (green) merge faster than their original \texttt{O} resolution counterparts. Increasing the resolution leads to a less efficient tidal stripping by the galactic environment and causes the clusters to retain more mass. In both systems 3 and 12, the \texttt{HI} models sink about 20\% faster. The \texttt{LOWM} model (red line) shrinks to the largest hard binary radius as the NSCs are about an order of magnitude less massive than the \texttt{FRe} model. }
    \label{fig:dr_dt_all}
     \end{center}
\end{figure*}

Our simulations all start roughly when the MBHs are $\Delta r \approx 300$ pc apart. We present visualizations of the sinking process for two of our models, \texttt{sys2\_FRe\_O} and \texttt{sys6\_FRe\_O}, in Figure \ref{fig:system_graphics}. In \texttt{sys2\_FRe\_O}, we find that the process is extremely efficient with the MBHs forming a bound binary and sinking to $\sim 10^{-3}$ pc within 16 Myr from the start of the simulation. The fast sinking is aided by the high stellar density of the surrounding bulge and the large mass of the NSC hosting $M_2$ with $M_{\rm{NSC},2} \approx 7.2 \times 10^6 M_{\odot}$. In fact, \texttt{sys2\_FRe\_O} is the fastest sinking model amongst all of the models simulated. This is in contrast to  \texttt{sys6\_FRe\_O}  where the MBHs take 85 Myr from the start of the simulation to form a bound binary. The NSC hosting $M_2$ is only $17\%$ as massive as that in \texttt{sys2\_FRe\_O} model. The central stellar density is also $10\times$ lower than that in \texttt{sys2\_FRe\_O} reducing the efficiency of DF leading to a longer inspiral.  

We examine the evolution of $\Delta r$ as a function of time in Figure \ref{fig:dr_dt_all} for all models, finding that the MBHs are able to sink and form a hard binary, shrinking significantly below $r_{\rm h}$. In all but one model, \texttt{sys1\_FRe\_O}, the MBHs sink within 100 Myr from the start of our simulations. \texttt{sys1\_FRe\_O} is unique in that it takes $\approx 300$ Myr for the MBHs to sink to below the hard binary radius. This is primarily caused due to the fact that the NSC hosting the secondary is only $1.5 \times 10^6 M_{\odot}$ and the initial separation of the MBHs is slightly larger than that in other simulations. Incidentally, \citet{Zhou2024MagicsII} find that system 1 also ends up stalling at the largest scale ($\gtrsim 400$ pc) amongst all systems studied in \citetalias{Zhou2024MagicsII}. 

While the sinking time is dependent on the galactic environment and the masses of the MBHs, it is still informative to examine the mean sinking time $\tau_{\rm sink,avg}$ across all our simulations. To calculate $\tau_{\rm sink}$ for each model, we note the time when the MBHs reach $\Delta r = 0.1$ pc, close to the average of influence radii across all systems. Using only the data from the \texttt{FRe} models to be consistent, we find that, $\tau_{\rm sink, avg} \approx 540$ Myr. We compare this to the average sinking time for the MBHs derived from \citetalias{Chen2024} suite and find that the our sinking times are, on average, 20 \% shorter. However, we emphasize that the the sinking time in the \citetalias{Chen2024} suite is calculated when the MBHs reach $\Delta r \approx 20$ pc. Thus the actual sinking time without the NSCs may be substantially higher. This is corroborated by comparing our models to those from the \citetalias{Zhou2024MagicsII} suite where the authors follow the evolution down to sub-pc scale and find sinking times of the order of $\sim$Gyr in simulations where the binary does not stall.

\begin{figure}
    \begin{center}
    \includegraphics[width=0.5\textwidth]{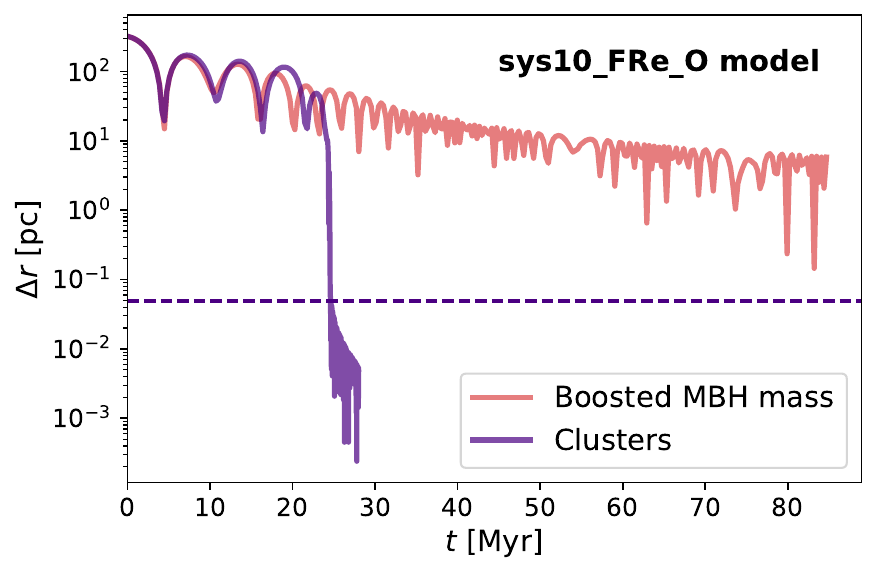}
    \caption{The binary separation $\Delta r$  as a function of time $t$ for \texttt{sys10\_FRe\_O} model with MBHs in NSCs (purple) and with the MBH masses boosted by the amount of mass present in the respective NSCs (red). While the early evolution is similar in both cases, indicating that it is dominated by DF from the increased mass, the same cannot be said for $t > 20$ Myr. The orbits shrink by roughly the same amount until the MBHs reach $\Delta r \sim 50$ pc, after which the tidal interactions between the NSCs drive the MBHs to rapidly sink to the center of the merged NSC in $< 0.5$ Myr and form a hard binary. However, the boosted MBH mass model does not demonstrate this rapid shrinking leading to a much longer sinking timescale. This indicates that NSCs are a key ingredient of rapid MBH binary formation.}
    \label{fig:clusters_vs_boosted}
     \end{center}
\end{figure}

One might ask whether this accelerated sinking is caused due to the additional mass of the NSCs surrounding the MBHs which enhances the DF force experienced by the MBH+NSC system or if it is due to the tidal interactions between the NSCs and the galactic environment. To examine this, we perform an experiment with \texttt{sys10\_FRe\_O} where, instead of adding the NSC surrounding the MBHs, we boost the mass of the MBHs by the same amount and plot the evolution of the MBH separation in Figure \ref{fig:clusters_vs_boosted}. We find that when the MBHs are separated by $\Delta r \gtrsim 100$ pc, the orbits shrink by roughly the same amount. There are only minor differences between the orbits in the first 20 Myr. This indicates the primary phenomenon driving the orbital shrinkage in this case is DF from the additional mass. However, once the MBHs reach a separation of $\approx 50$ pc, there is a period of rapid orbital shrinkage when the MBHs are surrounded by NSCs. This is driven by the tidal interactions between the two NSCs which exert torques on the MBHs leading to a rapid decrease in their angular momentum and energy. The MBHs form a hard binary at the end of this phase, which is very rapid and lasts $\leq 1$ Myr. This is consistent with the findings of \citet{Ogiya2020MNRAS.493.3676O} where the authors note that tidal interactions dominate over DF at separations of $\Delta r \sim 50$ pc. In the boosted MBH mass case the binary continues to shrink due to DF and is only able to reach  $\Delta r = 10$ pc in 85 Myr. This indicates the importance of extended stellar systems such as NSCs in accelerating the formation of bound MBHs. We also perform an additional simulation where we only embed only the secondary MBH in system 12 in an NSC and find that the MBHs do not form a binary within $\gtrsim 200$ Myr.

\texttt{sys12\_LOWM} takes substantially longer to sink than its more massive counterpart, \texttt{sys12\_FRe\_O} . This is caused due to the low mass of the NSCs, almost a factor of 10 smaller than that in \texttt{sys12\_FRe\_O}. Nevertheless, a hard binary forms around $\approx 500$ Myr. The binary settles to separation of $\lesssim 1$ pc and continues further hardening via three-body interactions. We compare the sinking time between this model and the equivalent system from the \citetalias{Zhou2024MagicsII} and find that they are similar. \citet{Zhou2024MagicsII} find that the MBHs sink to $\sim$ 1 pc around 600 Myr from the beginning of their simulation. The similarity in the sinking separations and timescales indicates the consistency between our simulations.

\subsection{Density profile of the formed NSC}

\begin{figure*}
    \begin{center}
    \includegraphics[width=1.0\textwidth]{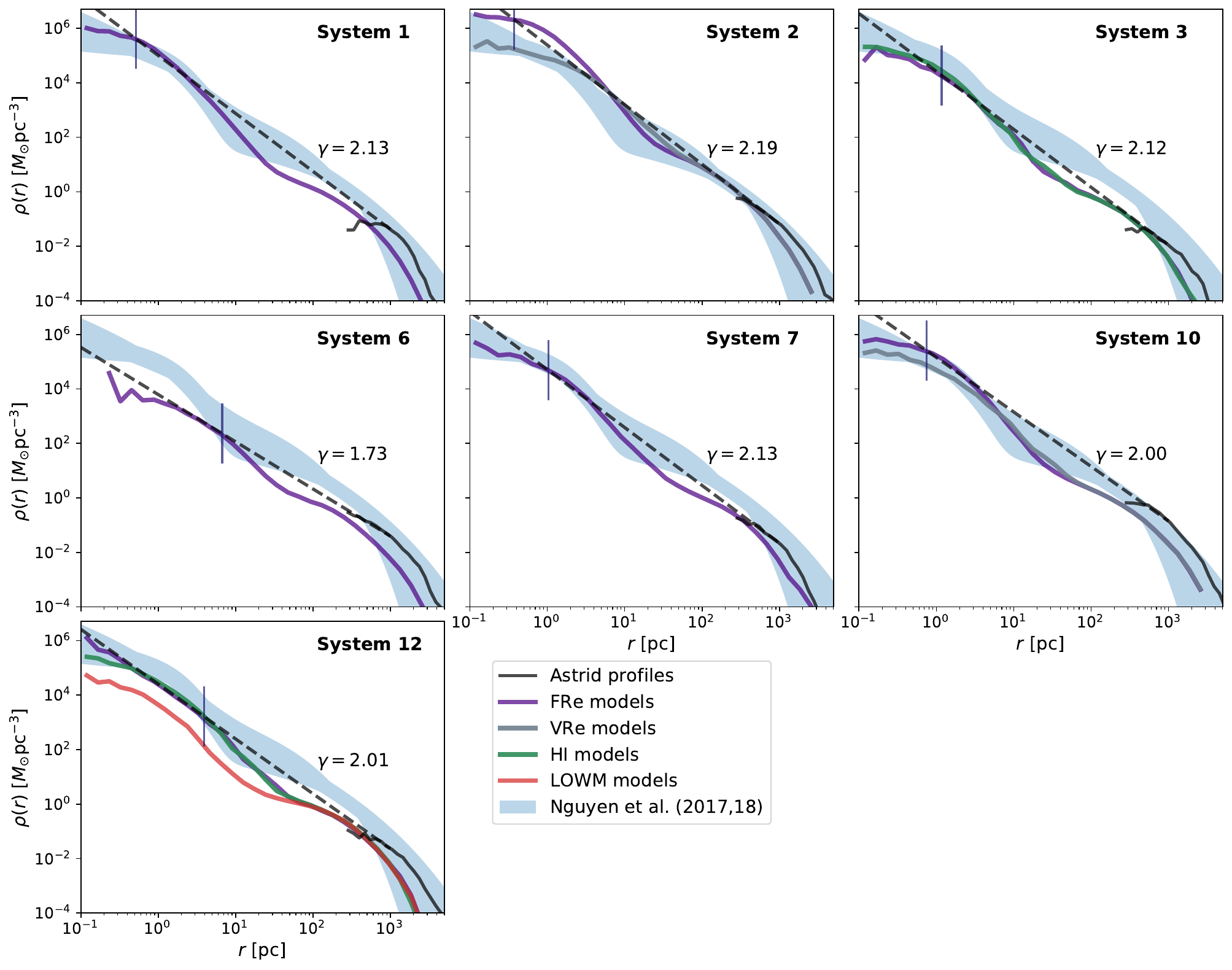}
    \caption{The stellar density $\rho(r)$ at the moment the NSCs merge as a function of distance from the center of potential of the system $r$ for our models and the equivalent systems from \texttt{ASTRID} (black line). The different line colors indicate the different models, similar to Figure \ref{fig:dr_dt_all}. The profiles are measured before any significant scouring effects due to the formation of the hard MBH binary. The \texttt{ASTRID} stellar profiles are more dense in the outskirts than our profiles as we neglect further galaxy mergers. The stellar density in $r < 10$ pc is quite consistent with observations (blue shaded region) of nucleated local dwarfs with similar galactic stellar masses \citep[][]{Nguyen2017ApJ...836..237N,Nguyen2018ApJ...858..118N} as the galaxies in our suite. We also calculate the slope extrapolation parameter $\gamma$, and the extrapolated stellar density profile (black dashed line), extrapolating the \texttt{ASTRID} density profiles from 1 kpc to the influence radius of the binary (vertical line). Averaging across simulations, we find $\gamma \approx 2.04$, with denser NSCs having a slightly higher $\gamma \approx 2.1-2.2$. }
    \label{fig:density_all}
     \end{center}
\end{figure*}

Once the two NSCs have merged and before a hard binary has formed, we measure the overall stellar density profiles and compare them to observations of known NSCs. We choose the set from \citet{Nguyen2017ApJ...836..237N,Nguyen2018ApJ...858..118N}, which includes density profiles of M32, NGC5102, NGC5206, NGC404, and NGC205, as well as sets from \citet{Pechetti2020ApJ...900...32P} and \citet{Georgiev2016MNRAS.457.2122G}. All these sets contain local NSCs, while our galaxies are present in the high redshift universe. Nevertheless, these comparisons allow us to understand how our models relate to present-day NSCs in similar mass galaxies. We first provide a comparison of our formed NSCs against the \citet{Nguyen2017ApJ...836..237N,Nguyen2018ApJ...858..118N} set as the galaxy stellar masses used in that set are comparable to those in our models.

Comparing density profiles across our models to the set from \citet{Nguyen2017ApJ...836..237N,Nguyen2018ApJ...858..118N} in Figure \ref{fig:density_all}, we find that stellar density in the inner 10 pc is quite consistent with that from \citet{Nguyen2017ApJ...836..237N,Nguyen2018ApJ...858..118N}. 
For \texttt{sys2\_FRe\_O} , we notice $\rho_{\rm{1 \: pc}} \approx 2\times 10^6 M_{\odot} \rm{pc}^{-3}$, highest among all our models and about a factor of 10 larger than the densest cluster from \citet{Nguyen2017ApJ...836..237N,Nguyen2018ApJ...858..118N}, M32. This is caused due to the high initial density of the NSC harboring the primary MBH. However, we notice that the \texttt{VRe} counterpart for the same system produces a density profile that is in agreement with the observations over the entire radii. \texttt{sys6\_FRe\_O} produces the least dense cusp among all our models (other than the \texttt{LOWM} model), with $\rho_{\rm{1 \: pc}} \approx 4\times10^{3} M_{\odot} \rm{pc}^{-3}$. This is unsurprising given the low initial masses and densities of the NSCs in that particular case.  

The densities in the \texttt{VRe} models are systematically lower than their \texttt{FRe} counterparts in the inner parsec, although they are higher in the outer parts at $r\gtrsim 10$ pc. This is caused due to the fact that the \texttt{VRe} models have a lower central density. Since the masses of the NSCs are not changed between the \texttt{FRe} and \texttt{VRe} models, they have higher densities in the outskirts. Our higher resolution NSC models, the \texttt{HI} models, produce profiles that are in strong agreement with their original resolution \texttt{O}  counterparts. We compare the density profiles at various radii between the \texttt{O} and \texttt{HI} models and find that the profiles differ by a factor of 2-3 at $r \sim 1$ pc at most. The \texttt{HI} models retain more mass than their \texttt{O} counterparts, leading to a slightly higher density in certain parts.

We do notice that our stellar bulge density to NSC density is somewhat larger than the observations from \citet{Nguyen2017ApJ...836..237N,Nguyen2018ApJ...858..118N}. This is primarily caused by our initial NSC generating prescription which produces slightly more massive NSCs compared to the galaxy stellar mass. However, core scouring caused by the MBH binary can lead to significant erosion \citep[e.g.,][]{Merritt2006ApJ...648..890M,Merritt2007ApJ...671...53M}. As we will show in the next section, it can lead up to a 70\% decrease in the density profile at $r=1$ pc which can produce density profiles that are more consistent with observations. 

In typical cosmological simulations, stellar profiles can only be reliably measured up to approximately 1 kpc. Post-processing studies often extrapolate the stellar density profiles using a fixed power law to calculate the density at the influence radius, $\rho_{\rm infl}$, of the binary \citep[][]{Chen2022MNRAS.514.2220C}. This is crucial because the hardening timescales are sensitive to $\rho_{\rm infl}$. The extrapolated profiles are defined by the slope parameter $\gamma$, which determines how the density $\rho_{\rm extra} \propto r^{-\gamma}$ behaves. For clarity, we note that the extrapolated $\gamma$ present here is different from  $\gamma_{\rm cl}$, the inner slope of the initial density profile of the clusters. Typically, $\gamma$ values between 1.5 and 2 are used, producing a wide range in the binary's hardening timescales. Here, we investigate the $\gamma$ parameter obtained from our simulations, which resolve the density down to sub-pc scales. We use the original stellar profiles from \texttt{ASTRID} for the same systems and measure $\gamma$ based on $\rho_{\rm infl}$ from our \texttt{SR} models and $\rho_{\rm 1 \; kpc}$ calculated from \texttt{ASTRID}. Examining $\gamma$ and the extrapolated density profiles in Figure \ref{fig:density_all}, we find that for most systems $\gamma \gtrsim 2.0$. For system 6, we find a lower gamma on account of two factors: the \texttt{ASTRID} profile is denser at $1$ kpc since the galaxy undergoes more mergers in the full volume simulation, and the NSC is less dense compared to NSCs in other systems. The galaxies in system 6 undergo more mergers than other systems as the MBHs are recorded as merging at a later snapshot in \texttt{ASTRID} compared to the other systems. 
This results in a lower $\gamma=1.73$ extrapolated profile. Taking the average across all systems, we find $\gamma_{\rm avg} = 2.04$. For post-processing in cosmological simulations, we recommend using this value to account for the effects of dense nuclear structures.

We also compare the stellar density in our models at $r=5$ pc with the $\rho-M_{\rm{*,gal}}$ relationship provided in \citet{Pechetti2020ApJ...900...32P} derived from a volume-limited sample of 27 galaxies. According to \citet{Pechetti2020ApJ...900...32P},  
\begin{equation} \label{equation:pechetti_5}
    \rm{log} (\rho_{\rm{5 \, pc}}) = (0.61 \pm 0.18) \left ( \frac{M_{*,gal}}{10^9 M_{\odot}} \right ) + (2.78 \pm 1.68).
\end{equation}
 Using the $M_{\rm *,gal}$ values from our models, we find that $\frac{\rho_{\rm sim}}{\rho_{\rm obs}} \sim 5$ when $M_{\rm *,gal} \leq 5\times10^8 M_{\odot}$ but increases to $\sim 20$ for $M_{\rm *,gal} \sim 10^9 M_{\odot}$. However, our values are within the limits of the uncertainty in the \citet{Pechetti2020ApJ...900...32P} relation. In future studies, we plan on investigating the effect of choosing $M_{\rm NSC}$ more conservatively to try and better replicate the observed relationship from our simulations. \texttt{sys12\_LOWM} produces a density profile that is lower than that calculated using equation \ref{equation:pechetti_5} by a factor of 3. This indicates that a realistic NSC mass lie in between our \texttt{O} and \texttt{LOWM} models. 

\begin{figure}
    \begin{center}
    \includegraphics[width=0.5\textwidth]{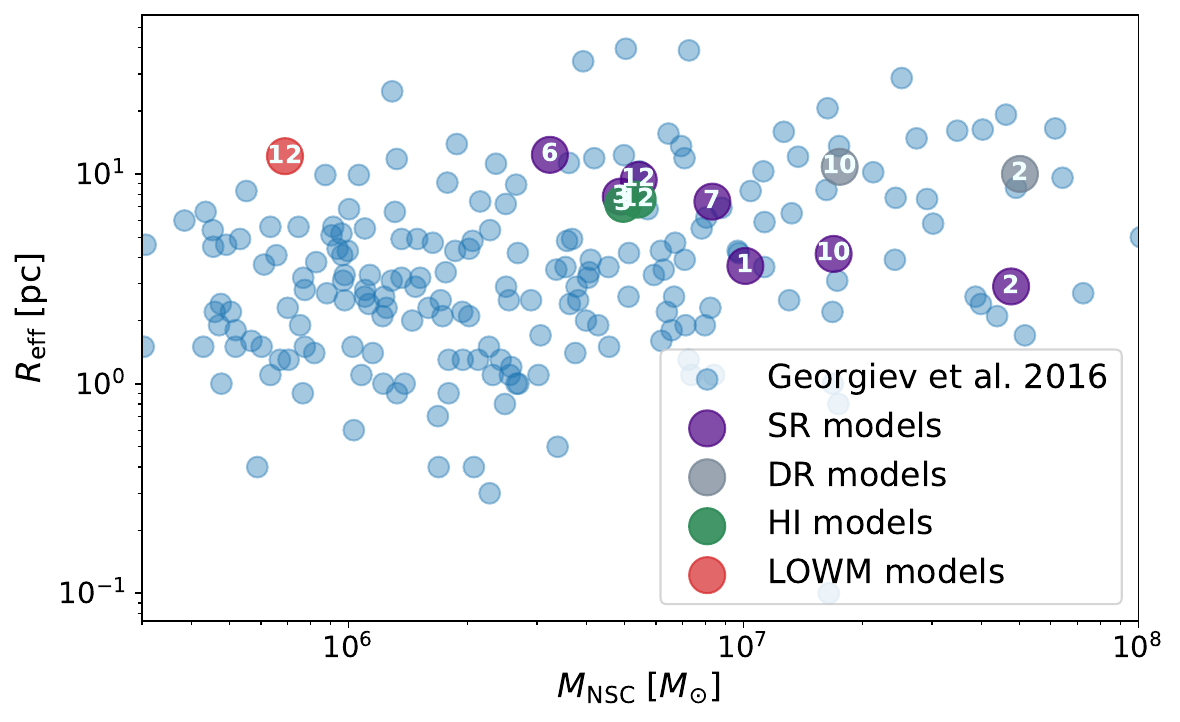}
    \caption{The effective radius of the final NSC $R_{\rm eff}$ as a function of the mass of the NSC $M_{\rm NSC}$ upon the merger of the two individual NSCs. The blue circles represent values obtained from \citet{Georgiev2016MNRAS.457.2122G} while the larger circles are values obtained from our simulations, with the color indicating the model type (similar to those from previous figures), and the numbers indicating the system number. We find that $R_{\rm eff}$ of the merged NSCs are quite consistent with those from observations. For the \texttt{FRe} models (purple circles), there is an inverse correlation between $R_{\rm eff}$ and $M_{\rm NSC}$. Thus the NSCs with $M_{\rm NSC}\geq 10^7 M_{\odot}$ are somewhat denser than the ones that are observed. In the \texttt{VRe} models (grey circles), we find that $R_{\rm eff}$ of the formed NSC is about a factor of 5 larger and more consistent with observations. We find that the effect of resolution is subdominant with minor differences in $R_{\rm eff}$ between the original resolution \texttt{O} and higher resolution \texttt{HI} models (green circles). For the \texttt{LOWM} model (red circle), we find that the merged NSC has $R_{\rm eff}$ that is about an order of magnitude larger than NSCs with similar masses. This is a result of the large initial $R_{\rm eff}$, larger than what we expect from observations. }
    \label{fig:reff_observations}
     \end{center}
\end{figure}

We compare the effective radii $R_{\rm eff}$ of our formed NSCs to those from \citet{Georgiev2016MNRAS.457.2122G} in Figure \ref{fig:reff_observations}. The simulated NSCs across various models match the observations. For the \texttt{FRe} models, we find that $R_{\rm eff}$ is negatively correlated with the mass of the merged NSC $M_{\rm NSC}$. For $M_{\rm NSC} \leq 10^7 M_{\odot}$, we produce NSCs are are slightly underdense than observed ones whereas for $M_{\rm NSC} \geq 10^7 M_{\odot}$, we produce NSCs that are denser than the ones observed. Nonetheless, our simulations produce NSCs with $R_{\rm eff}$ that lie within the limits of the observations. The \texttt{VRe} counterparts for systems 2 and 10 produce NSCs are on average more consistent with the observed data. While they are less dense than their \texttt{FRe} counterparts, they still sink in roughly the same amount of time. This indicates that despite the lower density, NSCs are still efficient at accelerating MBH mergers. The higher resolution \texttt{HI} models produce NSCs that have very similar $R_{\rm eff}$ as their \texttt{O} counterparts indicating that our results are convergent and robust.

\subsection{Core scouring}

\begin{figure*}
    \begin{center}
    \includegraphics[width=1.0\textwidth]{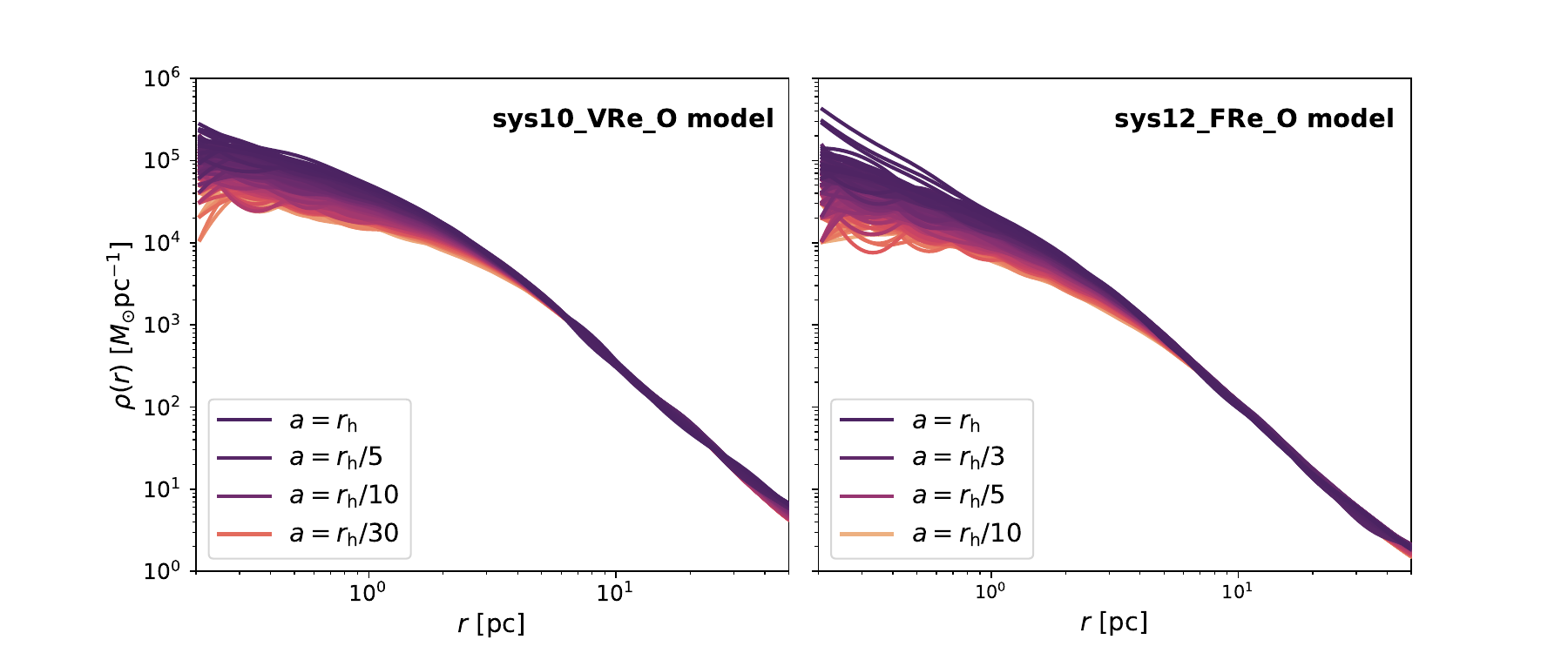}
    \caption{Evolution of the stellar density $\rho(r)$ as a function of the distance from the MBH binary $r$ over time as the MBH binary hardens. The binary hardens by ejecting particles from the cusp leading to the slow scouring of the initial cusp. Left: evolution of the density profile in the \texttt{sys10\_VRe\_O} model where the binary is evolved for a total duration of $26$ Myr in the hard binary phase. Right: same but for the \texttt{sys12\_FRe\_O} model and the evolution is followed for 33 Myr. The mass of the binary in this model is larger than that in \texttt{sys10\_VRe\_O} while the NSC mass is lower. This leads to a slower hardening and a larger erosion of the cusp.}
    \label{fig:core_scouring}
     \end{center}
\end{figure*}

The formation of a hard binary and the subsequent hardening process can lead to a substantial decrease in the central density of the formed cluster \citep[e.g.,][]{Merritt2006ApJ...648..890M,Merritt2007ApJ...671...53M}. The binary undergoes complicated scattering events with the stellar matter surrounding it leading to a complete or partial ejection of the particle from the system. This leads to a mass deficit in the center of the formed cluster. In order to examine how the density profile evolves as a function of the binary's semi-major axis $a$, we plot the density profile for \texttt{sys10\_VRe\_O} and \texttt{sys12\_FRe\_O} at several times in Figure \ref{fig:core_scouring}. These models were chosen since they had been evolved for sufficiently long time in the hard binary phase. For the \texttt{sys10\_VRe\_O} model, we find that the formation of the binary and subsequent hardening to $a = r_h/30$ leads to a significant decrease in the central stellar density profile. At $r=1$ pc, the scouring effect of the binary leads to a $5\times$ decrease in the density. A qualitatively similar observation is made for the \texttt{sys12\_FRe\_O} model. However, we find that the scouring of the density profile is more prominent here as a $\approx 7 \times$ decrease in the density profile is observed by the time the binary hardens to $a=r_{\rm h} / 10$. \texttt{sys12\_FRe\_O}, in fact, shrinks more slowly than the \texttt{sys10\_VRe\_O} model caused due to the larger binary mass and lower density of the NSC. This implies that a larger amount of mass needs to be ejected for the binary to shrink by the same amount as that in  \texttt{sys10\_FRe\_O} model leading to a more significant erosion of the stellar density within the same amount of time. 

\begin{figure}
    \begin{center}
    \includegraphics[width=0.5\textwidth]{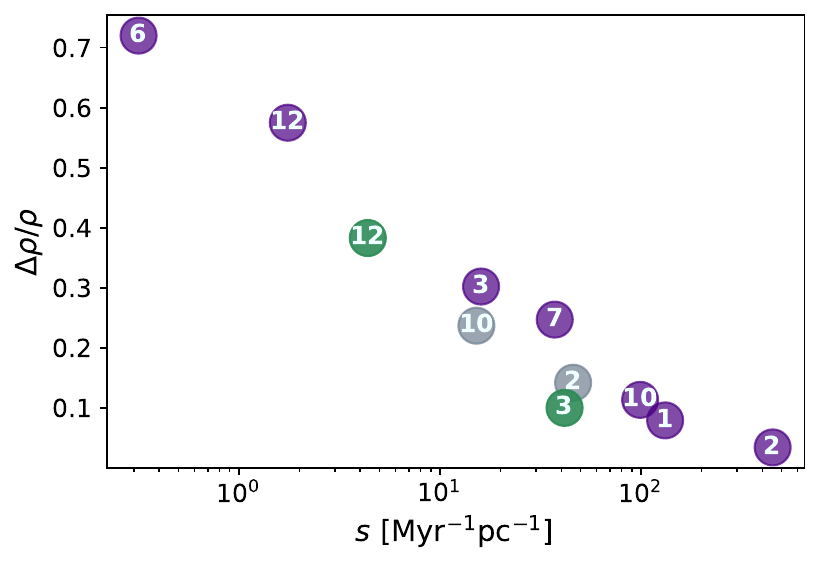}
    \caption{Relative change in the density $\Delta \rho / \rho$ at $r=1$ pc as a function of the binary hardening rate $s$. The circle number and color indicate the system number and model type respectively, as in Figure \ref{fig:reff_observations}. The values are measured when the binary has hardened to $a=r_{\rm h}/5$. There is an inverse correlation between  $\Delta \rho / \rho$ and $s$. A faster hardening implies a slower rate of erosion of the cusp as the initial cusp density is higher and/or the mass of the binary is lower, similar to findings from \citet{Merritt2007ApJ...671...53M}.}
    \label{fig:delta_rho_over_rho}
     \end{center}
\end{figure}

To understand how core scouring varies across all of our models, we compute the relative change in density $\Delta \rho / \rho$ computed at $r=1$ pc as a function of the binary hardening rate $s$. $\Delta \rho$ is computed by taking the difference between the initial density profile and that when the binary has hardened to $r_{\rm h}/5$ across all the models. The hardening rate $s$ is defined as 
\begin{equation}
    s \equiv \frac{d}{dt} \left ( \frac{1}{a} \right).
\end{equation}
We calculate $s$ from our simulations by fitting straight lines to the evolution of the inverse semi-major axis $1/a$ as a function of time every 0.3 Myr after the formation of the hard binary and then taking the mean value.  Plotting $\Delta \rho / \rho$ as a function of $s$ in Figure \ref{fig:delta_rho_over_rho}, we find that, a lower hardening rate correlates to a larger relative change in the density profile. For our slowest shrinking model a $\gtrsim 70\%$ decrease in density is observed whereas for the fastest shrinking model, it only leads to a change of $\lesssim 10\%$. This is qualitatively similar to the findings from \citet{Merritt2007ApJ...671...53M} where the authors find that the relative rate of mass loss is inversely proportional to the hardening rate. While our slower shrinking systems form a large flat core by the time they shrink to GW dominated stage, whereas for the denser systems, a cusp, albeit shallower, might still remain. Core scouring can also play a role in lowering the density such that our simulated systems match observations better. For example, for the \texttt{sys12\_FRe\_O} model, we observe that $\rho_{\rm{5 \, pc}}$ reduces by a factor of 2 by the time the binary has hardened to $r_{\rm h}/10$. This reduction in density leads to more consistency between the observed \citet{Pechetti2020ApJ...900...32P} relation and our simulated systems. 

One can also calculate the total mass deficit by taking the difference between the final and initial density profiles and then integrating over some radius. We follow this procedure for a variety of radii and calculate the mass deficit $M_{\rm def}$ by finding where the curves peak. To be consistent among models, we calculate the mass deficit when the binaries have hardened to $a_{\rm h}/5$. We find that the deficit is quite independent of the density profile used. For the bulk of our models we find that $M_{\rm def} \approx 1.4 \times M_{\rm bin}$ while for one of our models it peaks at $1.81 M_{\rm bin}$. This is qualitatively similar to the results from \citet{Khan2021MNRAS.508.1174K} although they find higher $M_{\rm def}$ in their models. Incidentally, similar to \citet{Khan2021MNRAS.508.1174K} we find that the $M_{\rm def}$ does not peak at $r_{\rm infl}$ but rather at distances greater than $r_{\rm infl}$. This indicates that stars beyond $r_{\rm infl}$ also contribute significantly to the hardening process, necessitating the inclusion of the bulge and DM halo.

\subsection{Evolution of the hard binary}

After the formation of a hard binary, we follow the evolution down to $a \lesssim r_{\rm h}/10$ for all models other than the \texttt{LOWM} model.  In this stage, the binary hardens via three-body interactions with the surrounding material. Under the full-loss cone approximation, we expect the MBH binary to harden at a fixed rate. Accordingly, we determine the hardening rate $s \equiv \frac{d}{dt} \left ( \frac{1}{a} \right )$ as mentioned in the previous section and study its evolution over time. We find that the hardening rate remains roughly constant, as expected, indicating that the binary loss-cone is filled by particles on centrophillic orbits. This is primarily caused due to the non-spherical geometry of the merged product which produces a torquing effect, leading to replenishment of the loss-cone. 

\citet{Quinlan1996NewA....1...35Q,Sesana2006ApJ...651..392S} provide a description of the energy loss during the hardening process. The hardening rate can be represented as 
\begin{equation} \label{equation:sesana_hardening}
    \frac{d}{dt} \left ( \frac{1}{a} \right ) = \frac{G H \rho}{\sigma}
\end{equation}
where $\sigma$ is the velocity dispersion, and $H$ is a dimensionless hardening parameter that is almost constant once the binary is sufficiently hard. \citet{Sesana2006ApJ...651..392S} performed three-body scattering experiments to numerically determine the hardening rate and found that $H \approx 16-20$. To determine the value of $H$ from our simulations, we measure $\rho$ and $\sigma$ at the influence radius $r_{\rm infl}$ of the binary as done in \citet{Sesana2015MNRAS.454L..66S}. We find that $H \approx 10$ for all our simulations. This is about 40\% lower than the theoretically predicted value of $16$ but consistent with the findings from other $N$-body simulations \citep[e.g.,][]{Fastidio2024arXiv240602710F}. This implies that although loss-cone interactions drive the binary towards GW stage, the binary does not harden in the full loss-cone regime \citep[][]{Vasiliev2019MNRAS.482.1525V}. We also compare the effect of resolution on the the value of $H$ by comparing the calculated values from the \texttt{O} and the \texttt{HI} models. For \texttt{sys3\_FRe\_O}, we find that $H=9.03$ whereas for \texttt{sys3\_FRe\_HI}, we find $H=9.48$, only about 5\% larger. Similarly, for \texttt{sys12\_FRe\_O}, we find $H=11$ while for \texttt{sys12\_FRe\_HI} we find $H=12.9$, about 17\% larger. This indicates that the dimensionless hardening parameter is only weakly sensitive to the resolution and is convergent in our \texttt{FRe} models. We do note that slight variations between $\rho$ and $\sigma$ among the \texttt{O} and \texttt{HI} models are present which affect the overall hardening rate $s$. Owing to the slightly higher density in the \texttt{HI} models, we find that $s$ is larger by a factor of 2 in those models compared to the \texttt{O} models.

\begin{figure*}
    \begin{center}
    \includegraphics[width=1.0\textwidth]{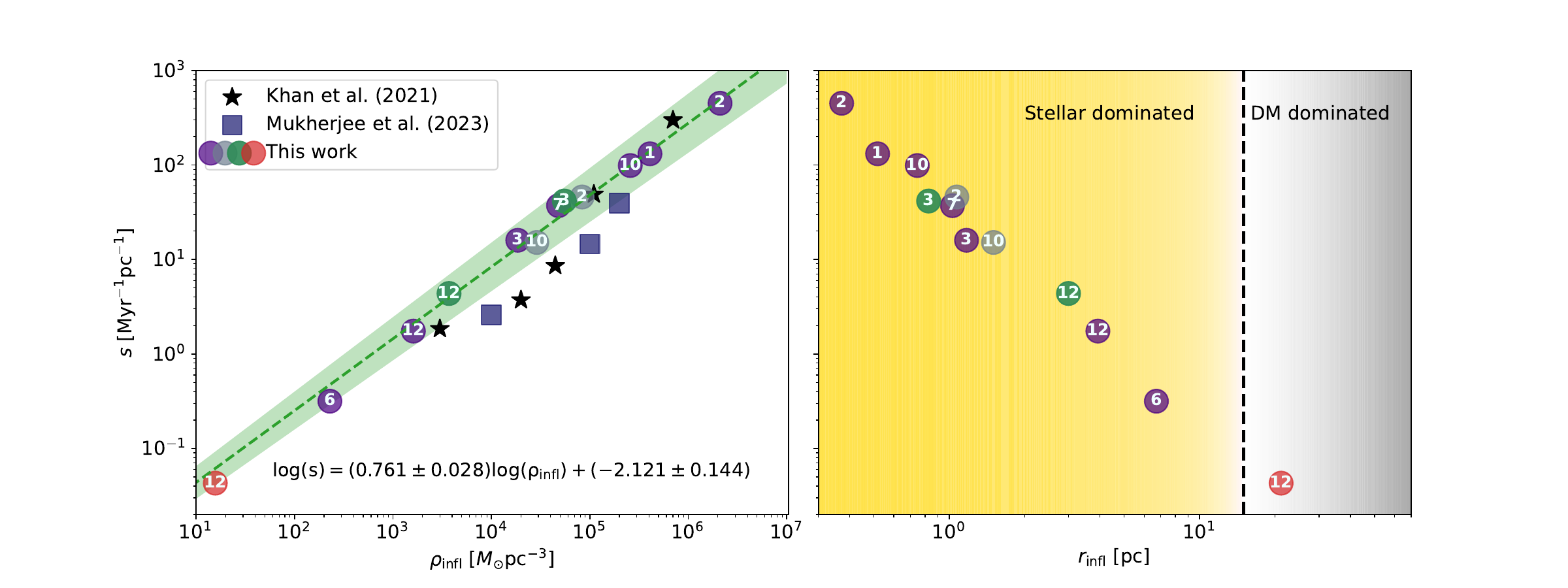}
    \caption{Left: The hardening rate $s$ as a function of the density at influence radius $\rho_{\rm infl}$. We find that $s$ is strongly correlated to $\rho_{\rm infl}$ and can be described by a power law (green line). The shaded region represents the uncertainty of our fit. Theoretically, we expect $s$ to be a linear function of $\rho_{\rm infl} / \sigma_{\rm infl}$. In our simulations $\sigma_{\rm infl}$ does not change a lot across our models, allowing us to obtain a one-parameter fit between $s$ and $\rho_{\rm infl}$. The obtained values are consistent with  those obtained from previous simulations with NSCs including \citet{Khan2021MNRAS.508.1174K} and \citet{Mukherjee2023MNRAS.518.4801M} differences of factors of 2-3 at most.  Right: $s$ but as a function of the influence radius $r_{\rm infl}$. Here we observe a negative correlation between $s$ and $r_{\rm infl}$ which is consistent with our expectations as a smaller $r_{\rm infl}$ implies a higher $\rho_{\rm infl}$.  We also find that in all but one case, the binary hardens primarily via three-body interactions with stars from the NSC and the bulge. However, for \texttt{sys12\_LOWM}, $r_{\rm infl}\approx 25.2$ pc and we find that DM contributes twice as much as the stars in the hardening rate. The black dashed vertical line represents where the contribution due to DM and stars is equal. }
    \label{fig:rhoinfl_vs_s}
     \end{center}
\end{figure*}

Given the diverse environments in our simulations, we aim to understand how the hardening rate $s$ compares across different models. In Figure \ref{fig:rhoinfl_vs_s} we compare $s$ as a function of $\rho_{\rm infl}$ and find a strong correlation between the hardening rate and the density at the influence radius across all our models. The model \texttt{sys2\_FRe\_O} hardens the fastest with $s=450.8 \rm{Myr}^{-1} \rm{pc}^{-1}$ which is unsurprising given its $\rho_{\rm infl}=2.1\times10^6 M_{\odot} \rm{pc}^{-3}$. Similarly, we find $s=0.04 \rm{Myr}^{-1} \rm{pc}^{-1}$ for \texttt{sys12\_LOWM} given the very $\rho_{\rm infl}=11 M_{\odot} \rm{pc}^{-3}$. \citet{Zhou2024MagicsII} find a hardening rate of $\approx 0.02 \rm{Myr}^{-1} {pc}^{-1}$ for system 12 from their simulations. Despite substantial methodological differences, the similarity in hardening rates between the two sets of simulations underscores the robustness of our results. Recently \citet{Khan2024arXiv240814541K} performed simulations of IMBH binary mergers in non-nucleated dwarfs. Their \texttt{D1.5C} model is has similar $\rho_{\rm infl}$ as our \texttt{sys12\_LOWM} model and we find  hardening rates comparable to their results. 

For models with intermediate $\rho_{\rm infl}$ such as \texttt{sys7\_FRe\_O}, $s=37.1 \rm{Myr}^{-1} \rm{pc}^{-1}$. We find that the relationship between $s$ and $\rho_{\rm infl}$ can be described by a simple power-law. We fit the values obtained from our simulations and find the relationship to be
\begin{equation} \label{equation:fitted_hardening}
    \begin{split}
    \rm{log}_{10} \left ( \frac{s}{\rm{Myr^{-1}} \rm{pc}^{-1}} \right) = (0.761 \pm 0.028) \rm{log}_{10} \left (\frac{\rho_{\rm{infl}}}{M_{\odot} \rm{pc}^{-3}} \right) \\ + (-2.121 \pm 0.144).
    \end{split}
\end{equation}
Interestingly, \citet{Khan2024arXiv240814541K} in simulations of IMBH binaries in non-nucleated dwarfs, find a linear relationship between $\rho_{\rm infl}$ and $s$ in the low $\rho_{\rm infl}$ regime. 
We additionally compare our values to those obtained from previous simulations with NSCs and find that for similar $\rho_{\rm infl}$, our hardening rates are consistent with those obtained from previous works. \citet{Khan2021MNRAS.508.1174K} study the hardening of IMBH binaries in a variety of nucleated dwarf galaxies, taking into account the effect of the bulge and NSC. We find that for $\rho_{\rm infl} \geq 10^5 M_{\odot} \rm{pc}^{-3}$, the hardening rates predicted by our simulations are fully in agreement with those from \citet{Khan2021MNRAS.508.1174K}. For example, the authors find that $s=49 \rm{Myr}^{-1}\rm{pc}^{-1}$ for NGC 404 where $\rho_{\rm infl}=1.1 \times10^5 M_{\odot} \rm{pc}^{-3}$. For similar $\rho_{\rm infl}$ such as in \texttt{sys2\_VRe\_O} where $\rho_{\rm infl}=8.3\times10^4 M_{\odot} \rm{pc}^{-3}$, we find $s=45.9 \rm{Myr}^{-1} \rm{pc}^{-1}$. At lower densities, there are differences of at most factors of $2-4$. For NGC 5206 \citet{Khan2021MNRAS.508.1174K} find $\rho_{\rm infl}=2\times10^4 M_{\odot} \rm{pc}^{-3}$ and $s=3.73 \rm{Myr}^{-1} \rm{pc}^{-1}$, whereas in \texttt{sys3\_FRe\_O}, we obtain $s=15 \rm{Myr}^{-1} \rm{pc}^{-1}$ for a similar $\rho_{\rm infl}$. These differences may be caused due to the differences in initial conditions and resolution. We also compare our results to those from \citet{Mukherjee2023MNRAS.518.4801M} where the authors studied the effect of MBH mergers in mass-segregated NSCs and find that our results are quite consistent especially for higher $\rho_{\rm infl}$. The differences at lower $\rho_{\rm infl}$ are, again, at most factors of 2-3. These could be caused due to differences in initial conditions, or due to the lack of a stellar bulge in \citet{Mukherjee2023MNRAS.518.4801M} as \citet{Khan2021MNRAS.508.1174K} showed that a non-trivial amount of the hardening is caused to the bulge stars, especially for equal mass ratio binaries. Theoretically we expect a linear relationship between $s$ and $\frac{\rho_{\rm infl}}{\sigma_{\rm infl}}$. We fit $s$ versus $\frac{\rho_{\rm infl}}{\sigma_{\rm infl}}$ and find that $s \propto \left (\frac{\rho_{\rm infl}}{\sigma_{\rm infl}} \right )^{0.96 \pm 0.04}$ in agreement with theory. 

Since we have both stars and DM in our simulations, an interesting question one might ask is which component contributes more to hardening. Examining Figure \ref{fig:rhoinfl_vs_s} where we compare the $s$ as a function of the influence radius $r_{\rm infl}$, we find that most of our systems harden by scattering with stars. The hardening rate is negatively correlated to $r_{\rm infl}$ implying that binaries with large $s$ harden by scattering with particles closer to the binary. Examining where $M_{\rm def}$ peaks along with $r_{\rm infl}$ values, we find that stars mainly from the cluster contribute more to the hardening of the binary for systems with large $s$. However, in particularly low density environments such as \texttt{sys6\_FRe\_O} and \texttt{sys12\_LOWM}, the situation is different. For the latter, we find that $r_{\rm infl} \approx 25.2$ pc. Here $\rho_{\rm DM,infl} = 11 M_{\odot} \rm{pc^{-3}}$ while $\rho_{\rm *,infl} = 2 M_{\odot} \rm{pc}^{-3}$. Comparing the hardening rates of different components using equation \ref{equation:sesana_hardening}, we find that the DM contributes $2\times$ as much to the hardening rate as stars. Therefore, in DM dominated dwarf galaxies lacking dense NSCs, the effect of the DM halo cannot be neglected as the binary primarily hardens by scattering with DM, consistent with the findings from \citet{Partmann2023arXiv231008079P}. Therefore, we expect DM to dominate the hardening in non-nucleated dwarfs and the effect of the DM halo cannot be neglected. The addition of the DM halo could potentially alleviate the issue of non-mergers of IMBH binaries in non-nucleated dwarf galaxies as noticed by \citet{Khan2024arXiv240814541K}. 

We also study the evolution of the eccentricity during this stage. The MBHs are initially on highly eccentric orbits given that \citet{Chen2024} find that a combination of high eccentricity and central stellar density is essential for the MBHs to sink sufficiently. Before the MBHs become bound, the orbital eccentricity is measured as 
\begin{equation} \label{equation:unbound_ecc}
    e = \frac{r_{\rm apo} - r_{\rm peri}}{r_{\rm apo} + r_{\rm peri}}
\end{equation}
where $r_{\rm apo}$ and $r_{\rm peri}$ are the relative apocenter and pericenter distances. Initially, the MBH pairs have unbound orbital eccentricities of $0.75-0.95$ across all systems. The mean eccentricity initially is 0.86 with a standard deviation of 0.06. During the DF phase, we notice a systematic decrease in the eccentricity of the pairs, similar to \citet{Gualandris2022MNRAS.511.4753G}. By the time the MBHs are 30 pc apart, the mean eccentricity reduces to 0.67.  This is caused due to the the circularizing effect of DF \citep{Gualandris2022MNRAS.511.4753G}. However, we notice that this phase introduces a lot of scatter in our eccentricity values, increasing the standard deviation to 0.2. This is unsurprising given stochasticity introduced due to our finite resolution. \citet{Nasim2020MNRAS.497..739N} find that this scatter is introduced by the merger process and affects binaries with $e>0.9$ more substantially. This makes inferring bound binary eccentricities as a function of the initial eccentricity challenging. 

\begin{figure}
    \begin{center}
    \includegraphics[width=0.5\textwidth]{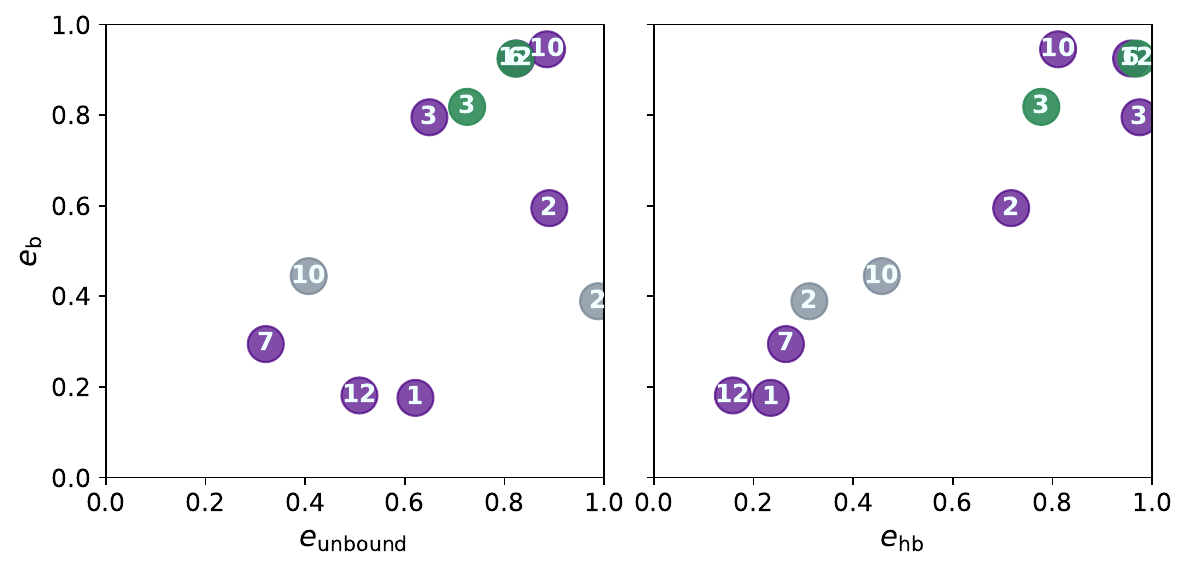}
    \caption{Evolution of eccentricity across all our models. Left: The bound eccentricity $e_{\rm b}$ as a function of the unbound orbital eccentricity as calculated using equation \ref{equation:unbound_ecc} of the pair measured at a separation of $\Delta r=30$ pc. $e_b$ is measured before the black holes form a hard binary and when they are separated by a distance $r_{\rm bound}$ as defined in equation \ref{equation:gualandris_eb}. Although the initial orbital eccentricity of all the black holes are 0.85-0.95, $e_{\rm unbound}$ shows a wide scatter with the mean around 0.75. $e_{\rm b}$ is correlated to $e_{\rm unbound}$ but stochasticity associated with higher $e_{\rm unbound}$ produces a wide variety of $e_b$. Right: $e_b$ as a function of the eccentricity at the hard binary separation $e_{\rm hb}$. We notice that a higher $e_b$ implies a higher $e_{\rm hb}$. During the hardening stage, we find \texttt{sys3\_FRe\_O}, \texttt{sys12\_FRe\_HI} and \texttt{sys6\_FRe\_O} are able to achieve almost radial orbits. The differences among \texttt{HI} and \texttt{O} models persist in this stage with \texttt{sys12\_FRe\_O} containing a binary with an eccentricity of 0.15 while that in \texttt{sys12\_FRe\_HI} is 0.95.  }
    \label{fig:eunb_eb_ehb}
     \end{center}
\end{figure}

To understand how the eccentricity of the bound binary behaves as a function of the unbound eccentricity, we measure the unbound eccentricity $e_{\rm unbound}$ when the MBHs are separated by 30 pc and compare it to the bound eccentricity $e_b$ in Figure \ref{fig:eunb_eb_ehb}. The latter is measured according to equation 6 from \citep{Gualandris2022MNRAS.511.4753G} where the eccentricity is measured at a separation $r_{\rm bound}$ where 
\begin{equation} \label{equation:gualandris_eb}
    r_{\rm bound} = M_{\rm enc} \left(0.1(M_p + M_s) \right).
\end{equation} For the former, we choose to measure the unbound eccentricity at $\Delta r = 30$ pc as we find that the initial eccentricity produces a wide range of $e_b$ erasing any correlations. Similar findings were also noted by \citet{Fastidio2024arXiv240602710F} where the authors found a wide range of $e_b$ when the initial eccentricity $e_0$ was greater than 0.9. Examining Figure \ref{fig:eunb_eb_ehb}, we notice that $e_{\rm unbound}$ is correlated to $e_b$ but there is significant scatter in $e_b$ when the $e_{\rm unbound}$ is high. In \texttt{sys10\_FRe\_O} we find both $e_{\rm unbound}$ and $e_b$ are greater than 0.9, the situation for \texttt{sys2\_FRe\_O} and \texttt{sys2\_VRe\_O} are different where the pair has $e_{\rm unbound} \approx 0.9$ but $e_b$ is significantly lower. We also note that there are differences between the formed binary among our \texttt{O} and \texttt{HI} models. While both \texttt{sys3\_FRe\_O} and \texttt{sys3\_FRe\_HI} have similar $e_{\rm unbound}$ and $e_{b}$, the same cannot be said for the \texttt{sys12\_FRe\_O} and \texttt{sys12\_FRe\_HI}. In \texttt{sys12\_FRe\_O}, the bound binary has a low eccentricity of 0.2 whereas in the \texttt{sys12\_FRe\_HI}, the bound binary has an eccentricity of 0.9. Our investigations reveal that  NSCs in \texttt{sys12\_FRe\_HI} merge somewhat earlier when the unbound orbital eccentricity is still large. In the \texttt{O} model, the pair undergoes circularization before the NSCs merge leading to a lower bound eccentricity. This is an unfortunate effect of stochasticity, especially for initially eccentric binaries, making inferences of bound binary eccentricity from the unbound eccentricity challenging , which is consistent with \citet[][]{Rawlings2023MNRAS.526.2688R}. On the other hand, examining the right panel of Figure \ref{fig:eunb_eb_ehb}, we find there is a tighter correlation between the hard binary eccentricity $e_{\rm hb}$ and $e_b$. For binaries with $e_b < 0.6$, we find that $e_{\rm hb}$ remains nearly the same as $e_b$, whereas for higher $e_b$, we notice a slight increase in the eccentricity. Longer evolution reveals that binaries with high eccentricities grow their eccentricity over time in the hard binary phase, leading faster mergers via the emission of GWs.

The evolution of eccentricity is also affected by the slope of the cusp in the original NSCs. \citet{Gualandris2022MNRAS.511.4753G} find that for denser cusps, the eccentricity of the formed binary is systematically lower. Similarly, due to collisional relaxation under a two-component mass species, \citet{Mukherjee2023MNRAS.518.4801M} find that relaxed systems which have denser cusps, form binaries with lower eccentricities. In our models, the density profile of the NSCs initially follows a shallow $\gamma=0.5$ cusp. Thus eccentricities of our formed binaries can be considered to be an upper limit. In future studies, we plan on varying the initial cusp of the NSCs, and/or incorporating a mass function to understand how that affects our results.

\section{Discussion} \label{sec:discussion}
\subsection{Gravitational wave timescales}

\begin{figure}
    \begin{center}
    \includegraphics[width=0.5\textwidth]{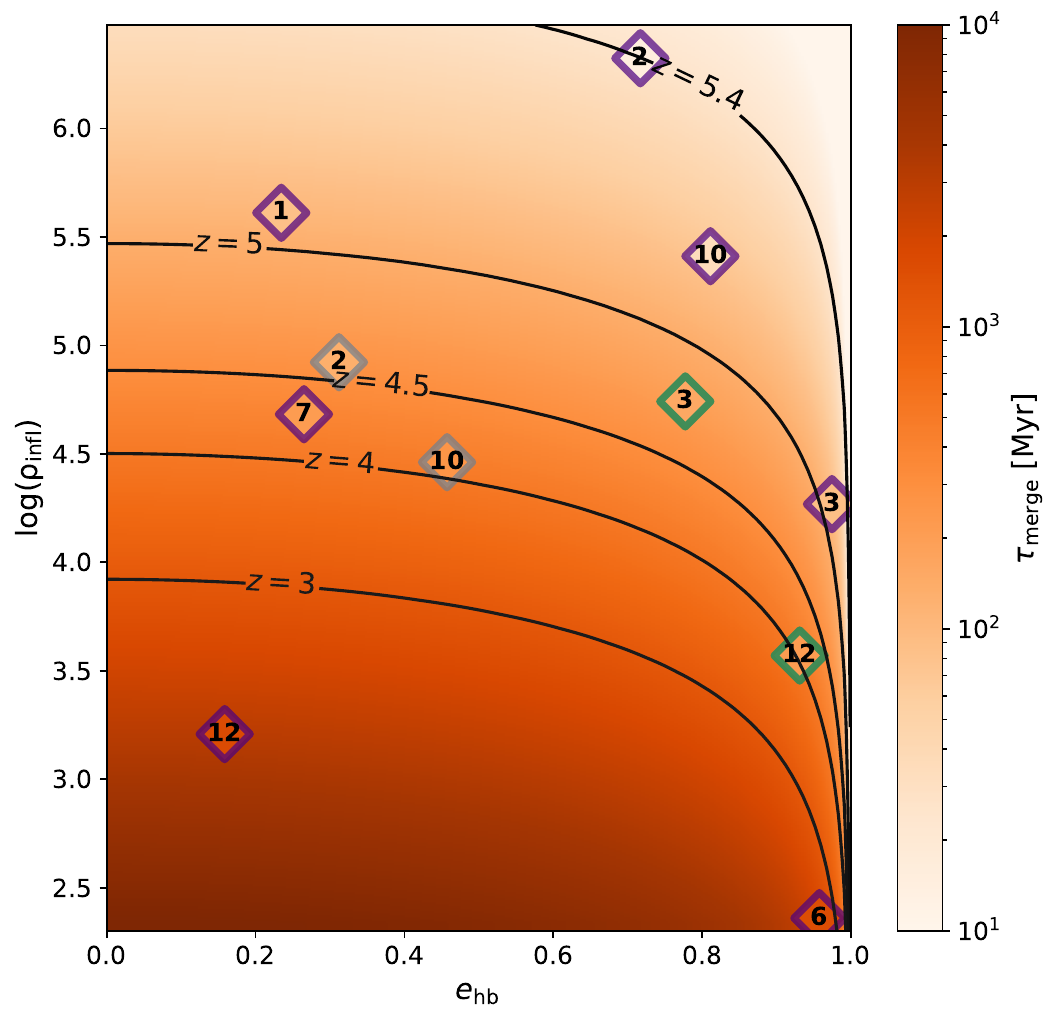}
    \caption{The merger timescale $\tau_{\rm merge}$ as a function of the density at influence radius $\rho_{\rm infl}$ and the eccentricity at hard binary radius $e_{\rm hb}$. The diamonds represent $\tau_{\rm merge}$ of our different models with the color and number indicating type of model and system number respectively, similar to that used in previous figures. Using $\rho_{\rm infl}$, we calculate the hardening rate $s$ using equation \ref{equation:fitted_hardening} which allows us to approximate the merger time following equations 13 - 18. The contours are drawn approximating that the galaxy merger process begins at $z=9$ and it takes 500 Myr for the MBHs to sink to hard binary radius. Under this approximation, we find that binaries that harden in environments with $\rho_{\rm infl} > 10^{4.5} M_{\odot} \rm{pc}^{-3}$ merge by $z=4$. Eccentric binaries merge faster and can lead to high redshift mergers even in somewhat lower density environments. Nevertheless, we find that seed black holes require a high stellar density environment to merge at high redshifts, suggesting that most mergers that happen at high redshifts occur in NSC dominated environments. We find in 8 our of the 12 models we simulate, the MBH binary merges before $z=4$. }
    \label{fig:gw_heatmap}
     \end{center}
\end{figure}

We use a semi-analytic method similar to that used in previous studies \citep[e.g.,][]{Gualandris2012ApJ...744...74G,Gualandris2022MNRAS.511.4753G} to calculate the GW merger timescale $\tau_{\rm merge}$ of our models. We assume that the hardening rate $s$ remains constant and that there is no change in eccentricity. Under this assumption, we can write the change in semi-major axis $a$ due to three-body hardening over time as 
\begin{equation}
    \frac{da}{dt} \bigg |_* = -s^2 a.
\end{equation}
Then, the overall evolution of $a$ can be written as follows:
\begin{equation}
    \frac{da}{dt} = \frac{da}{dt} \bigg|_{GW} + \frac{da}{dt} \bigg|_{*}
\end{equation}
We use \citet{Peters1964PhRv..136.1224P} equations to determine $\frac{da}{dt} \bigg|_{\rm GW}$ and $\frac{de}{dt} \bigg|_{\rm GW}$. According to \citet{Peters1964PhRv..136.1224P},
\begin{gather}
    \frac{da}{dt} \bigg|_{GW} = -\frac{64}{5}\beta \frac{F(e)}{a^{3}} \\
    \frac{de}{dt} \bigg|_{GW}  = -\frac{304}{15}\beta \frac{e G(e)}{a^{4}}
\end{gather}
where
\begin{gather}
    \beta = \frac{G^3}{c^5}\left(M_{p}M_{s}\left(M_{p} + M_{s}\right)\right) \\
    G(e) = (1-e^{2})^{-5/2}\left(1+\frac{121}{304}e^2\right) \\
    F(e) = (1-e^2)^{-7/2}\left(1+\frac{73}{24}e^2+\frac{37}{96}e^4\right) .
\end{gather}
We solve the set of coupled differential equations assuming that the initial semi-major axis and eccentricity for the differential equations are $a_{\rm hb}$ and $e_{\rm hb}$ respectively. 
Equation \ref{equation:fitted_hardening} allows us to to calculate $\tau_{\rm merge}$ as a function of $\rho_{\rm infl}$ by fixing $e_{\rm hb}$. For simplicity, we calculate $\tau_{\rm merge}$ as a function of $\rho_{\rm infl}$ and $e_{\rm hb}$ assuming $M_p = 10^5 M_{\odot}$ and $q=0.5$. To simplify our calculations, we estimate $\tau_{\rm merge}$ by starting from $a=0.02$ pc. This is roughly close to the mean of $a_{\rm hb}$ across the models considered in this study. We present  $\tau_{\rm merge}$ as a function of $\rho_{\rm infl}$ and $e_{\rm hb}$ under these set of assumptions in Figure \ref{fig:gw_heatmap} with merger timescales from our models overlaid. We note that $\tau_{\rm merge}$ is measured from the time the binary reaches $a_{\rm hb}$. We also draw contours assuming that the galaxies merge at $z=9$, and that the MBHs take $500$ Myr to sink to form a hard binary. Examining Figure \ref{fig:gw_heatmap} we find that mergers at $z>4$ require $\rho_{\rm infl} > 10^{4.3} M_{\odot} \rm{pc}^{-3}$ even for moderately eccentric binaries with $e_{\rm hb}=0.8$, indicating that high redshift MBH seed binaries with masses $\lesssim 2\times 10^5 M_{\odot}$ merge only in NSC dominated environments. Among our models, binaries embedded in high density NSCs such as \texttt{sys2\_FRe\_O} or those with $e_{\rm hb} > 0.9$ such as \texttt{sys12\_FRe\_HI} merge before $z=4$. One caveat of this calculation is that we do not consider the growth of eccentricity during the hard binary phase which may lead to faster mergers. Additionally, recent studies have shown that rotation in galactic nuclei can affect the eccentricity of the binary at formation \citep[][]{Rasskazov2017ApJ...837..135R}. Nuclei counter-rotating with respect to the binary can lead to an almost radial binary at formation accelerating merger timescales by factors up to 10 whereas nuclei co-rotating lead to systematically lower eccentricities \citep[e.g.,][]{Mirza2017MNRAS.470..940M}.   Nevertheless, our results strongly suggest that high-$z$ mergers happen in dense stellar environments that are only possible in NSCs. Additionally, our results coupled with those from \citetalias{Zhou2024MagicsII} suggest that naked seed MBHs do not even make it to the galactic center efficiently. \citet{Zhou2024MagicsII} find that in only one of the six systems simulated the MBHs seeds form a binary. This is attributed to the presence of the nuclei around both MBHs. Even a low mass NSC such as that present in our \texttt{LOWM} model helps the MBHs sink efficiently. Such stellar structures around the seeds were neglected in previous studies \citep[][]{Ma2021MNRAS.508.1973M,Partmann2023arXiv231008079P} and the authors found that the sinking process of seeds, especially low mass ones, is highly erratic. Our study in association with \citet{Zhou2024MagicsII} corroborates this but finds that in the presence of extended dense stellar systems such as NSCs, the seed sinking problem can be resolved. Low mass seeds are more likely to be formed in such high density environments indicating that future studies cannot neglect the effect of NSCs.

\subsection{Implications for future detection}

\begin{figure}
    \begin{center}
    \includegraphics[width=0.5\textwidth]{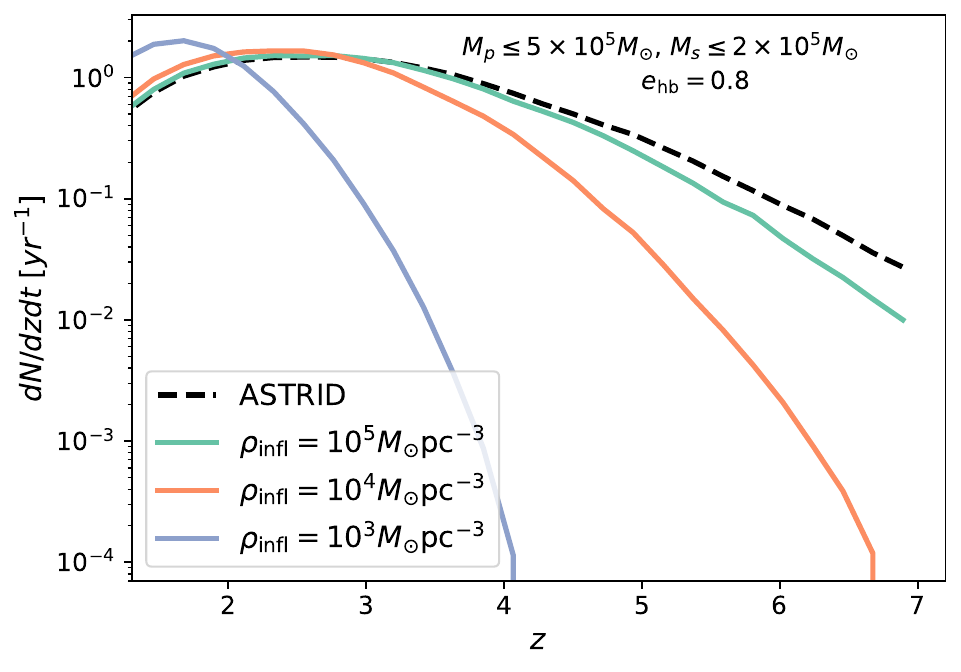}
    \caption{The merger rate per year $\frac{dN}{dz dt}$ as function of the redshift $z$. We extract the merger data from the \texttt{ASTRID} simulation for MBH pairs where $M_{p} \leq 5\times 10^5$ and $M_{s} \leq 2\times 10^5$, and add a hardening delay time based on different $\rho_{\rm infl}$ using equation \ref{equation:fitted_hardening} and the semi-analytic approach described in Section 6.1. For simplicity, we assume that the MBH binaries have hard binary eccentricity $e_{\rm hb} =0.8$. We find dense stellar environments dominate high-$z$ mergers at $z>4$ strongly suggesting that NSCs are the dominant channel for MBH seed mergers at high redshift. In bulge-like environments with $\rho_{\rm infl}=10^3 M_{\odot}$, the merger rate is much lower by factors of 600 compared to $\rho_{\rm infl}=10^5 M_{\odot}$ at $z=4$ and peaks at $z=1.67$.  Our calculation assumes that all mergers take place in nucleated environments, which may not hold in reality. However, for similar mass galaxies, \citet{Neumayer2020A&ARv..28....4N} report that 60\% may be nucleated. This would imply that NSC dominated evolution would be the primary channel for seed MBH mergers. }
    \label{fig:merger_rate_estimate}
     \end{center}
\end{figure}

What do our results imply for future detection? Using the results from our simulations, we can estimate the effect of different environments on the merger rate. We use the mergers recorded in the \texttt{ASTRID} simulation \citep{Ni-Astrid,Bird-Astrid} and select a subset where the MBH masses are: $M_p \leq 5.5\times 10^5 M_{\odot}$ and $M_s \leq 2\times10^5$. We record the redshift of the mergers and add a delay time based on the hardening timescale of the binary by calculating $s$ from $\rho_{\rm infl}$ using equation \ref{equation:fitted_hardening}. For simplicity we assume $e_{\rm hb} = 0.8$ and $a_{\rm hb} = 0.02$ pc. This calculation does not take into account the time delay due to DF as the mergers in \texttt{ASTRID} are recorded when the MBHs are $\sim$ kpc apart from each other. As noted in previous sections and in \citet{Zhou2024MagicsII}, the delay time due to DF dominated sinking can be of the order of $\sim 100$ Myr for MBH seeds surrounded by NSCs whereas for naked MBH seeds it is larger. Nevertheless, our calculation provides an upper limit on merger rate in a particular environment. Following \citet{Chen2022MNRAS.514.2220C}, once the delay time has been added, we calculate the merger rate as a function of the redshift $\frac{dN}{dzdt}$ as
\begin{equation}
    \frac{dN}{dz dt} = \frac{d^2 n(z)}{dz dV_c} \frac{dz}{dt} \frac{dV_c}{dz} \frac{1}{1+z}
\end{equation}
where $V_c$ is the comoving volume.
To calculate $\frac{d^2 n(z)}{dz dV_c}$ from the simulation,  we approximate it as 
\begin{equation}
    \frac{d^2 n(z)}{dz dV_c}  = \frac{N(z)}{\Delta z V_{\rm sim}}
\end{equation}
where $\Delta z$ is the width of the redshift bin, $N(z)$ is the total number of mergers in that particular bin, and $V_{\rm sim}$ is the volume of the simulation. For \texttt{ASTRID}, $V_{\rm sim} = (250 \rm{Mpc} / h)^3$. We add the delay time based on $\rho_{\rm infl}=10^3, 10^4, \rm{ and \,} 10^5 M_{\odot} \rm{pc}^{-3}$ and present the merger rate in Figure \ref{fig:merger_rate_estimate}. We find that at $z=4$, $\frac{dN}{dz dt} = 0.63 \rm{yr}^{-1}$ when $\rho_{\rm infl}=10^5 M_{\odot}$, $\frac{dN}{dz dt} = 0.33 \rm{yr}^{-1}$ when $\rho_{\rm infl}=10^4 M_{\odot}$, and $\frac{dN}{dz dt} = 10^{-3} \rm{yr}^{-1}$ when $\rho_{\rm infl}=10^3 M_{\odot}$. This suggests that most, if not all, detectable high redshift seed mergers happen in dense stellar environments, similar to the conclusions found by \citet{Khan2024arXiv240814541K}. In non-nucleated bulges stellar density is typically $\lesssim 10^3 M_{\odot}$ which results in three orders of magnitude fewer mergers at $z=4$. We also find that when $\rho_{\rm infl}=10^5 M_{\odot}$, the merger rate peaks at $z=2.76$ with a rate of $1.52 \rm{yr}^{-1}$, while for  $\rho_{\rm infl}=10^4 M_{\odot}$ the rate peaks at $z=2.33$ at a rate of $1.66 \rm{yr}^{-1}$. In low density environments such as $\rho_{\rm infl}$, we find that mergers peak later around $z=1.67$ at a rate of $2.01 \rm{yr}^{-1}$. Our calculations are approximate as we do not consider the fraction of galaxies that have nucleation and additional effects such as hierarchical three-body hardening which can also accelerate mergers. For the former, observations suggest anywhere between 20\% to 60\% of the galaxies with $M_{*} = 10^8 - 10^9 M_{\odot}$ may be nucleated \citep{Neumayer2020A&ARv..28....4N}. We will present a more detailed analysis of mergers taking into account the DF time delay, evolution of eccentricity, and nucleation fraction in a future work. 
Despite this, the implications of our work are robust and we find that the seed sinking problem can be resolved if the MBH seeds are embedded in NSCs. Additonally, our results suggest that high-redshift MBH seeds are an important source of GWs for future GW missions such as LISA which should be able to detect a few such mergers. This would be crucial to further constrain MBH seeding models.

\section{Conclusions} \label{sec:conclusions}
MBH seeds are one of the most important sources of gravitational waves (GWs) that will be detectable by future low-frequency GW detectors like LISA. The merger rates of MBH seeds at high redshifts can shed light on, and constrain seeding models. Previous studies have shown that the process by which MBH seeds sink to the centers of galaxies is inefficient, a phenomenon known as the seed-sinking problem. However, these analyses overlook the influence of extended stellar systems, such as nuclear star clusters (NSCs) surrounding the seeds. Combined with the findings of \citetalias{Zhou2024MagicsII}, which highlight the importance of these stellar systems in aiding the sinking process, our results strongly indicate that NSCs are a critical factor in accelerating both the sinking and merger of MBH seeds.

In this study, we utilize high-resolution N-body re-simulations of MBH seeds embedded in NSCs within high-redshift dwarf galaxies, employing up to $10^7$ particles using the FMM-based code \texttt{Taichi}. Our objective is to examine how NSCs influence the sinking and merger dynamics of MBH seed binaries, motivated by the findings of \citetalias{Zhou2024MagicsII}, which demonstrated that only MBH seeds retaining extended stellar systems around them are capable of sinking and forming binaries in high-$z$ dwarf galaxies. Building on \citetalias{Chen2024} and \citetalias{Zhou2024MagicsII}, we analyze the detailed dynamics of MBH seeds within NSCs orbiting the remnant host galaxy. For these models, we adopt a simple prescription for determining NSC masses, following \citetalias{Chen2024} and \citetalias{Zhou2024MagicsII}, where the total mass of both NSCs is set equal to the mass enclosed within the central 100 pc of the nucleus of the merged galaxy. The resulting NSCs in our simulations have masses ranging from $3\times10^5 M_{\odot}$ to $3\times10^7 M_{\odot}$ and are spherically symmetric, isotropic, and follow the \citet{Dehnen1993MNRAS.265..250D} density profile with a shallow cusp near the MBH.

%\tiziana{a more specific sentence about how the companion paper - btw please refer to it as MAGICII - motivated this study should be added here - I added below - please change/improve}
%we use the FMM-based code \texttt{Taichi} to simulate N-body models directly derived from high-resolution resimulations of cosmological galaxy mergers. %\tiziana{In MAGICSII, we found that MBHs that retain (/are not stripped) stellar structure can migrate effectively to the center and potentially lead to high-z MBH seed mergers. Here we assessed the detailed dynamical response of MBHs embedded in star clusters, orbiting in  a remnant host galaxy. }
%We systematically investigated the effect of NSCs on the sinking and merger timescales of MBH seeds in various cosmologically informed environments. 
%\tiziana{I would say something about how you set up the NSC --> information used from MAGICSII in terms of the mass of the clusters - of course your will vary that but it would be useful to remind people how you 'inform' your environment / what makes this a re-simulation}

The main conclusions of our work can be summarized as follows:
\begin{itemize}
    \item MBHs that are embedded in NSCs of all masses and sizes investigated in this study form a hard binary. NSCs
    assist MBH seeds in sinking to sub-pc scale. For more massive and denser NSCs, this process is extremely efficient, with MBHs sinking to $\lesssim 0.1$ pc within 15-20 Myr from the start of our simulation. The overall sinking times are, on average, 20\% faster than those found in \citetalias{Chen2024}. When the NSCs reach a separation of $\approx 50$ pc , we observe an accelerated decline in separation due to tidal interactions between the NSCs, leading to the formation of a hard MBH binary in $\lesssim 0.5$ Myr. Notably, this effect is absent when the MBH masses are boosted by the same amount, indicating that tidal effects of NSCs are crucial in helping MBHs form bound binaries. In the absence of NSCs surrounding either MBH seed, we find that the sinking process is inefficient. Our results converge upon increasing the resolution and we find that higher resolution models sink $10-20$\% faster.
    \item The density and effective radius of the central NSC produced after the merger are consistent with those of observed NSCs. Specifically, our density profiles fall within the range of known nucleated dwarfs drawn from \citet{Nguyen2017ApJ...836..237N,Nguyen2018ApJ...858..118N}. Additionally, the density of our NSCs aligns with the $\rho-M_{\rm *,gal}$ relationship from \citet{Pechetti2020ApJ...900...32P}. While the final effective radii of the merger product are somewhat smaller than expected for more massive NSCs in models using a fixed initial effective radius, our investigations reveal that selecting the initial effective radius based on \citet{Pechetti2020ApJ...900...32P} mitigates this issue. This paves the way for future studies to investigate more realistic models of NSCs derived from observational relations.
    \item We find that binaries harden efficiently when embedded in sufficiently dense NSCs. As expected from theoretical predictions, the hardening rate $s$ is proportional to the density at the influence radius $\rho_{\rm infl}$. In our densest model, $s\approx 450 \rm{Myr}^{-1} \rm{pc}^{-1}$, while in the least dense model, $s\approx 0.04 \rm{Myr}^{-1} \rm{pc}^{-1}$. Comparing our hardening rates to recent studies of MBH binaries in NSCs, we find consistent results, with differences of at most 2-4 when $\rho_{\rm infl}\sim 10^4 M_{\odot} \rm{pc}^{-3}$. Although three-body interactions with stars from the NSCs and bulge primarily drive the hardening, dark matter (DM) plays a dominant role in low-mass clusters. In our lowest mass model, the hardening rate due to DM is twice that due to stars. Based on our initial conditions, half of our models yield bound binaries with high eccentricity ($e>0.9$). Although the average unbound eccentricity decreases during the initial dynamical friction stage, substantial stochasticity makes predicting the bound binary eccentricity from the initial orbital eccentricity challenging.
    \item We follow the evolution of the binary into the GW merger stage using semi-analytic methods, assuming that the hardening rate $s$ remains constant and incorporating GW effects via the \citet{Peters1964PhRv..136.1224P} equations. Using the $s-\rho_{\rm infl}$ relationship derived in this work, we determine the merger timescales of seed binaries with $M_{\rm bin} \approx 1.5 \times 10^5 M_{\odot}$ in various environments. We find that a combination of high eccentricity and high stellar density is necessary for seeds to merge at high redshifts ($z > 4$). Even with a hard binary eccentricity of $\sim 0.9$ our models predict that $\rho_{\rm infl}$ should exceed $3\times10^4 M_{\odot} \rm{pc}^{-3}$, achievable only in nucleated environments. We find that 8 out of 12 of our models merge by $z=4$. Combined with the conclusions from \citet{Zhou2024MagicsII}, which indicate that naked MBH seeds do not sink efficiently to the center, our findings strongly suggest that NSCs are a crucial component in the seed sinking process. %\tiziana{the sentence above should be moved to the very beginning of the conclusions? 
    %Perhaps use this (or similar) instead of the sentence I added. It is a bit out of place in this rather specific bullet point}
    %also be careful to use MAGICII instead of Zhou}
    We also predict the merger rate of MBH seed binaries using data from \texttt{ASTRID} and add a hardening delay time based on different $\rho_{\rm infl}$, finding that detectable mergers at high redshifts would originate from sources embedded in dense nuclei. For $\rho_{\rm infl} = 10^3 M_{\odot} \rm{pc}^{-3}$, our results predict that the merger rate is 2-3 orders of magnitude lower than when $\rho_{\rm infl} \geq 10^4 M_{\odot} \rm{pc}^{-3}$ .
\end{itemize}

While our NSC models are idealized and do not account for gas effects, accretion, or MBH spin, this work lays the groundwork for more comprehensive future studies. Our findings highlight the critical role of NSCs in shaping the dynamics and evolution of seed MBHs, demonstrating that the seed sinking problem can be mitigated by the presence of extended stellar systems. With future integration of MBH spins, GW recoil, and enhancements to the FMM kernels, \texttt{Taichi} stands out as a powerful tool for addressing the complex problem of MBH mergers in galactic nuclei.

%% Also note that the akcnowlodgment environment does not support long amounts of text. If you have a lot of people and institutions to acknowledge, do not use this command. Instead, create a new \section{Acknowledgments}.
\begin{acknowledgments}
The authors thank Hy Trac for useful discussions. DM and UNDC acknowledge support from the McWilliams Center visitor program. DM acknowleges support from the Pittsburgh Supercomputing Center-McWilliams Center seed grant. We acknowledge the usage of the Vera computing cluster supported by the McWilliams Center and the Pittsburgh Supercomputing center.
DM thanks Nick Lewis for helpful discussions regarding image processing.
\end{acknowledgments}






\bibliography{references}{}

\bibliographystyle{aasjournal}



\end{document}